\documentclass[fleqn,usenatbib]{mnras}

\usepackage{newtxtext,newtxmath}
\usepackage[T1]{fontenc}
\usepackage{graphicx}	
\usepackage{amsmath}	

\title[The tails of NGC 752]{A tale of caution: the tails of NGC 752 are much longer than claimed}

\author[H.M.J. Boffin et al.]{
Henri M. J. Boffin$^{1}$\thanks{E-mail: hboffin@eso.org},
Tereza Jerabkova$^{2,1}$,
          Giacomo Beccari$^{1}$, and Long Wang$^{3,4,5}$
\\
$^{1}$European Southern Observatory, Karl-Schwarzschild-Strasse 2, 85748 Garching bei M\"unchen\\
$^{2}$European Space Agency (ESA), European Space Research and Technology Centre (ESTEC), Keplerlaan 1, 2201 AZ Noordwijk, The Netherlands\\
$^{3}$ School of Physics and Astronomy, Sun Yat-sen University, Daxue Road, Zhuhai, 519082, China \\
$^{4}$ Department of Astronomy, School of Science, The University of Tokyo, 7-3-1 Hongo, Bunkyo-ku, Tokyo, 113-0033, Japan\\
$^{5}$ RIKEN Center for Computational Science, 7-1-26 Minatojima-minami-machi, Chuo-ku, Kobe, Hyogo 650-0047, Japan
}
\date{Accepted 2022 May 24. Received 2022 May 24; in original form 2022 January 22}

\pubyear{2022}

\begin{document}
\label{firstpage}
\pagerange{\pageref{firstpage}--\pageref{lastpage}}
\maketitle

\begin{abstract}
Understanding the exact extent and content of tidal tails of open clusters provides useful clues on how field stars populate the Milky Way. We reanalyse, using \textit{Gaia} EDR3 data, the tails around the open cluster NGC 752. Compared to previous analyses, we look at a much wider region around the cluster and use first the convergent point method, coupled with a clustering analysis using \textsc{dbscan}.  We find that the cluster, located 433 pc away and well described by a Plummer profile, has very long and asymmetric tails, extending more than 260 pc on the sky (from tip to tip) -- four times larger than previously thought -- and contains twice as many stars. Numerical models computed with \textsc{petar} serve as a guide and confirm  our analysis. The tails follow the predictions from the models, but the trailing tail appears slightly distorted, possibly indicating that the cluster had a complicated history of galactic encounters. Applying an alternative method to the newly developed compact convergent point method, we potentially  trace the cluster's tidal tails to their full extent, covering several thousands of parsecs and more than 1\,000 stars. Our analysis therefore opens a  new window on the study of open clusters, whose potential will be fully unleashed with future \textit{Gaia} data releases.
\end{abstract}




\section{Gaia and tidal tails}
A majority of stars are  known to form in clusters embedded in molecular cloud cores \citep{2003ARA&A..41...57L,2022arXiv220106582D}, which expand after the expulsion of the residual gas to the radii of open clusters \citep{BK17}, in the process of which they can appear as associations when emerging as ensembles from their molecular clouds. Open clusters are particularly interesting as they provide great opportunities to study a single population of stars, characterised by a given age and metallicity. They thus serve, among others, as ideal probes of single and binary stellar evolution, over a very wide range of ages, chemical compositions, total masses, and locations in the Galaxy \citep[][]{2010RSPTA.368..755K}.  Such clusters  evolve with time, undergoing mass loss, relaxation, mass segregation, encounters with molecular clouds, and tidal disruptions, that will make them dissolve in the galactic field \citep{BK17,2019ARA&A..57..227K}. 

The tidal tails of stellar clusters are unique structures, whose study can reveal information on the initial conditions of cluster formation and the Galactic potential and its substructures \citep{2021A&A...647A.137J,Wang2021}. 
Stars escape an open cluster with relative velocities of a few km/s, thereby slowly drifting from the cluster and following very similar orbits. The cluster thus develops thin leading and trailing tidal tails, which are expected to be symmetric and S-shaped, depending on the orbit and physical properties of the cluster, as well as the Galactic potential \citep{Kuepper+10,Thomas+18}. Moreover, open clusters often have orbits that take them out of the Galactic mid-plane, and perturbations by the bar and spiral patterns of the Galaxy may lead to non-axisymmetric tails. 

Detecting the tidal tails around open clusters is very challenging for two reasons: their often relatively young ages means their tails won't be very large, while their low stellar content ($\approx 10^2-10^3$ stars) imply that they lost likely only a few hundreds of stars, which are often confined to the Galactic disc and thus heavily contaminated by field stars. 
Such detection is, however, now made possible thanks to  the ESA \textit{Gaia} satellite. Thus, following the release of the \textit{Gaia} DR2 catalogue, tidal tails have been found around nearby ($<300\,$pc) open clusters \citep[][and references therein]{2021A&A...647A.137J}, with ages between 100\, and 800\, Myr. 
With its higher precision in proper motions and parallaxes, the early data release 3 (EDR3) of \textit{Gaia} promises an even better detection of the tails. 

Here, we aim at revisiting, with the help of \textit{Gaia} EDR3, the tidal tails of the open cluster NGC 752. With an almost solar metallicity, an age between 1--2 Gyr, a very small amount of extinction, a galactic latitude of $-23$ degrees, and a distance of about 450 pc \citep{1991A&AS...87...69P,2018ApJ...862...33A}, NGC 752 is an ideal test case to probe the ability of \textit{Gaia} at detecting tidal tails, and an excellent complement to those clusters where tails were already detected.

\citet{2021MNRAS.505.1607B} recently used \textit{Gaia} EDR3 to identify the members of NGC 752 within 5 degrees around its center. The authors discovered the existence of the cluster's tidal tails, with an extent of about 35 pc on either side of the cluster.  We will show here that the actual tails of NGC 752 are in fact much longer, when retrieved with different methods. First, we will show the predictions for the properties of  tidal tails of an NGC 752-like cluster as obtained using numerical simulations. Then, we will apply the \textsc{\textsc{dbscan}} clustering algorithm in the convergent point framework. We will also make use of our numerical simulations to apply an alternative method, derived from the compact convergent point method. Finally, we will analyse the stellar content in the cluster and its tails, taking also into account binaries.

\begin{figure}
    \centering
    \includegraphics[width=\columnwidth]{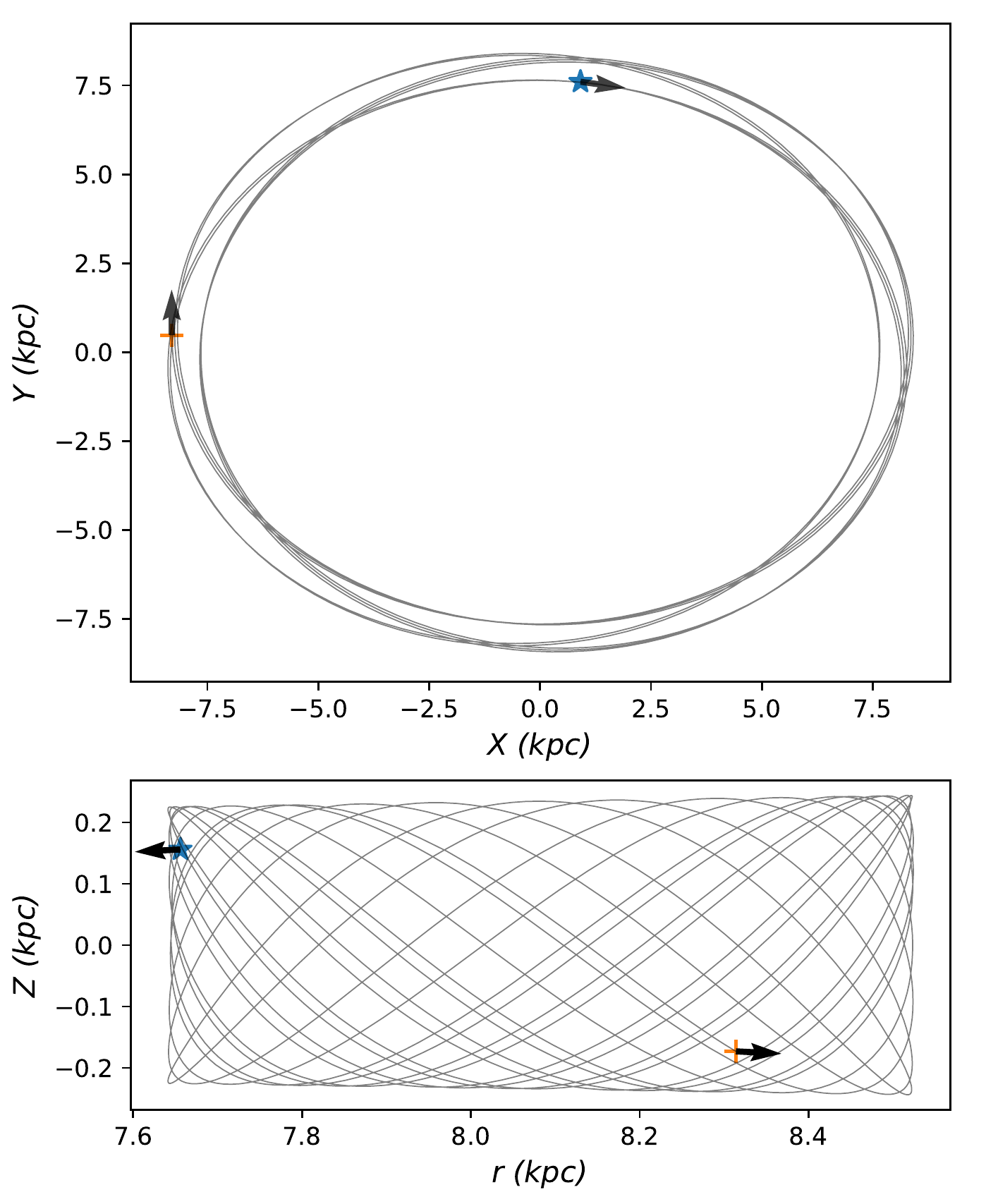}
    \caption{The orbit of the NGC 752-like cluster in Galactocentric coordinates. 
    The upper panel shows the orbit projected in the 
     $x-y$ plane, while the lower panel illustrates the orbit projected in the $r-Z$ plane, where $r$ is the
      radial distance projected onto the X-Y plane, that is, $r = \sqrt{X^2+Y^2}$. 
      The blue star and the orange cross represent the initial and the present-day positions, respectively. 
      }
    \label{fig:orbit}%
\end{figure}

\begin{figure*}
	\includegraphics[width=18cm]{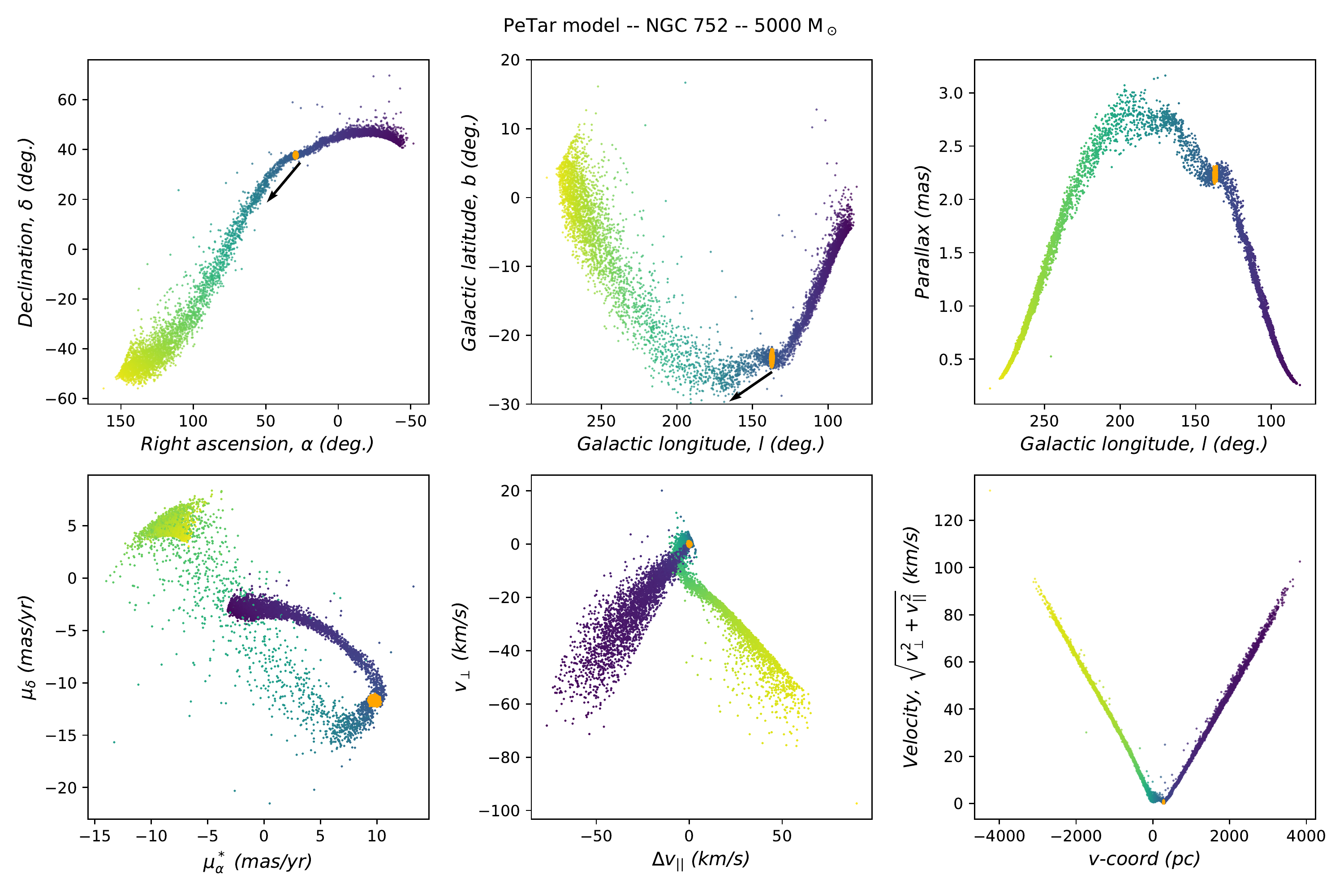}
    \caption{The outcome of the \textsc{petar} simulations of an NGC 752-like cluster and its tidal tails. The plots show from left to right and top to bottom: the stars in celestial coordinates, right ascension ($\alpha$) and declination ($\delta$); in Galactic longitude, $l$, and latitude, $b$; in the plane of Galactic longitudes and parallaxes; in proper motions; in convergent velocities; and in the plane of Cartesian galactic coordinate, $v$, and total convergent velocities. In each case, the colours correspond to the Galactic longitudes, while the central cluster is shown in orange. The vectors  show the direction of motion of the cluster.}
    \label{fig:petar}
\end{figure*}

\begin{figure*}
	\includegraphics[width=18cm]{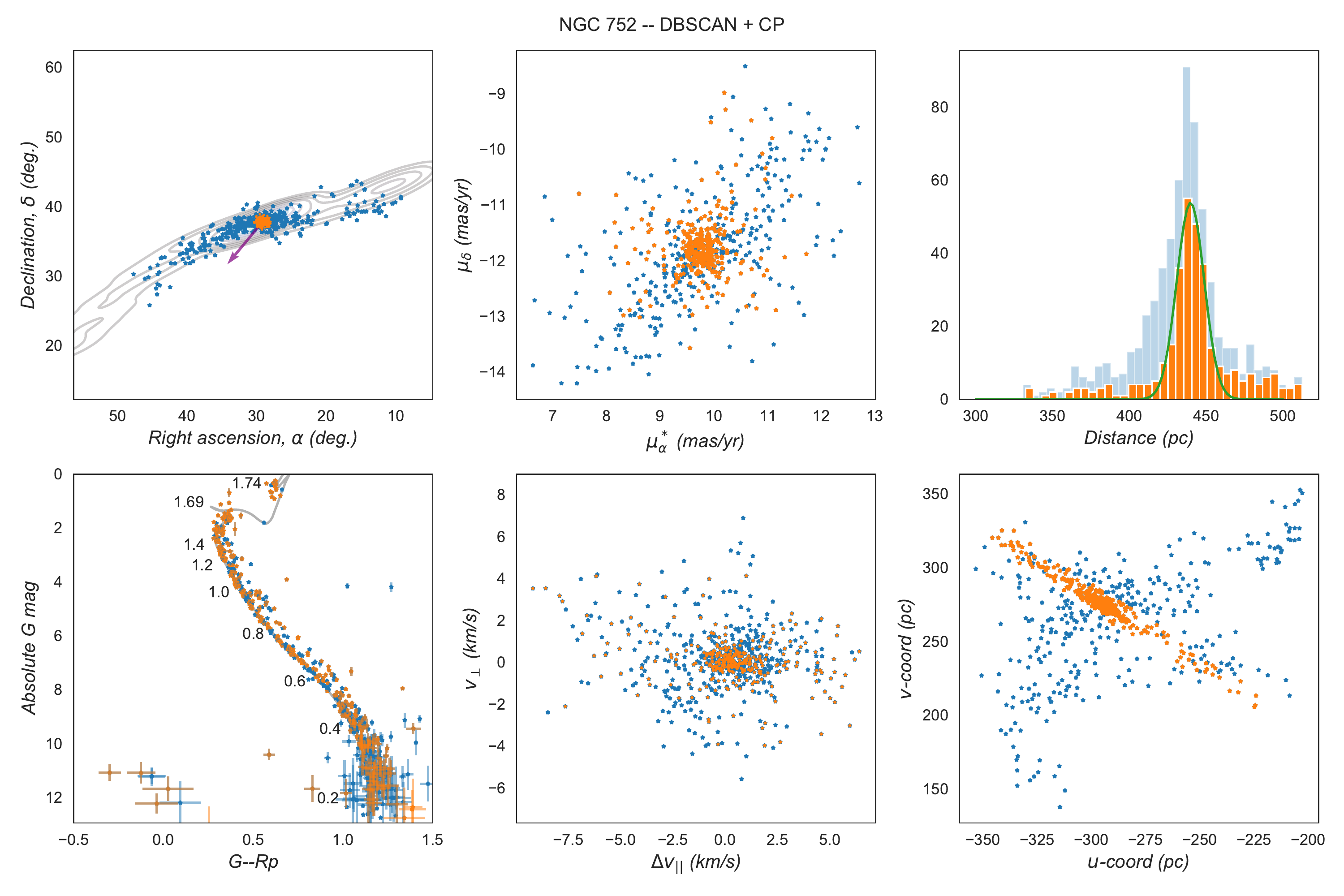}
    \caption{NGC 752 and its tidal tails as selected from \textit{Gaia} eDR3 data by \textsc{dbscan} in the 5D space of $v_\perp, \Delta v_{||}, u, v, w$. The plots show from left to right and top to bottom: the stars in celestial coordinates, right ascension and declination; in proper motions;  the distribution of the distances (in parsecs); in the \textit{Gaia} colour-magnitude diagramme (CMD); in the ($\Delta v_{||},v_\perp$) velocity space; and in the Cartesian galactic coordinates, $u-v$. In each cases, the colours correspond to either the full sample (blue) and the central parts of the cluster (orange). In the first panel, the positions of the stars in our numerical simulation are shown as the underlying contours. The vector shows the direction of motion of the cluster. In the CMD, a PARSEC isochrone of age 1.75 Gyr, [Fe/H]=$-0.13$ is shown, while the data have been corrected for interstellar extinction using \textsc{Bayestar2019}. The numbers correspond to the masses of the single stars at the given location (in solar masses).}
    \label{fig:tail}
\end{figure*}

\section{Petar models}
Having a $N$-body model with computed observable parameters proved to be an essential asset when searching/interpreting stellar clusters and their tidal tails \citep[][]{2021A&A...647A.137J,Wang2021}.
We use the $N$-body code \textsc{petar} \citep{Wang2020_Petar} to simulate the evolution of an NGC~752-like star cluster. 
\textsc{petar} uses the framework for developing particle simulation codes \citep[\textsc{fdps};][]{Iwasawa2016,iwasawa2020,namekata2018} for high performance,  and a slow-down algorithmic regularization \citep{Wang2020sdar} for handling dynamics of binaries and close encounters. The \textsc{sse} and \textsc{bse} \citep{Hurley2000,Hurley2002,Banerjee2020} codes are used for single and binary stellar evolution, while the \textsc{galpy} code with the \textsc{MWPotential2014} Galactic mass model is used for the Galactic potential \citep{Bovy2015}. 
The initial mass function (IMF) of \cite{2001MNRAS.322..231K}, with stars having masses above 0.08 M$_\odot$, is used, and no primordial binaries are considered here.

As in the case of the Hyades in \cite{2021A&A...647A.137J}, we use the present-day observed on-the-sky positions and velocities of the NGC~752 cluster and integrate backwards for 1.75 Gyr to obtain a realistic orbit in the Galactic potential.
Thus the initial conditions of the cluster in the simulation in Galactocentric coordinates is\\
$[X, Y, Z]$ = 0.912, 7.601, 0.157 kpc; \\
$[V_x, V_y, V_z]$ =  234.3,$-31.7$,  $-12.2$ km/s,\\
while the final values are\\
$[X, Y, Z]$ = $-8.293$, 0.273, $-0.144$ kpc;\\
$[V_x, V_y, V_z]$ = $-6.8$, 213.33,  $-13.4$ km/s.\\
In these coordinates, the Sun's position is \\
$[X , Y , Z ] = -8.000, 0.0, 0.015$ kpc;\\ 
and its velocity is\\
$[V_x, , V_y, , V_z, ] = 10.0, 235.0, 7.0$ km s$^{-1}$.

The orbit of the cluster is shown in these Galactocentric coordinates in Fig.~\ref{fig:orbit}, while the results of the simulations are shown in Fig.~\ref{fig:petar}, where it is obvious that the tails span a very wide area on the sky -- about 200 degrees from tip to tip. This corresponds to about five kpc! Given the age of the simulation, such an extent corresponds to drifting velocities of about 1.5--2 km/s.

It is important to notice that, by transporting the simulation into the realistic physical space, we can have an immediate idea of the possibility of using the Gaia catalogue to identify the cluster's tails. The plots in Fig.~\ref{fig:petar} indicate that the tails are covering a wide range of parallaxes, from about 0.3 to  3 mas, as well as making a very complicated pattern in proper motions. This already shows that traditional techniques that only look at clustering around the cluster's proper motions will miss a large part of the tails. The plot at the bottom-right, showing the convergent velocity (see below), $\sqrt{v_\perp^2 + \Delta v_{||}^2}$, as a function of the Galactic coordinate $v$ in the Galactic Cartesian system ($u, v, w)$, illustrates the use of the compact convergent method. 

The cluster was initially composed of stars for a total mass of 5\,000 M$_\odot$, distributed according to the above mentioned IMF. After the 1.75 Gyr of the simulation, all stars more massive than about 1.77 M$_\odot$ have ended their lives as compact objects, and the total mass was reduced to about 3\,200 M$_\odot$, of which 2\,800 M$_\odot$ are in normal stars (that is, not stellar remnants). Of these $\approx$ 860 M$_\odot$ (720 M$_\odot$ in normal stars) are within the central cluster itself, with the rest being in the tails. 

\section{Finding tails in Gaia}

\subsection{Using \textsc{dbscan}}
The result of \citet{2021MNRAS.505.1607B} was based on a study of a region of 5 degrees around NGC 752, which corresponds to about 40 pc at the distance of the cluster. As demonstrated by \citet{2021A&A...647A.137J} in the case of the Hyades, much longer tidal tails can be expected around open clusters, and this is even more true for NGC 752, which is three times older than the Hyades -- as confirmed by our simulations shown in the previous section. We have therefore decided to revisit this study, also using \textit{Gaia} EDR3, but considering a much larger circular region with a radius of 24 degrees -- corresponding to 200 pc --  around NGC 752, as an initial step. A query was made to the \textit{Gaia} archive to retrieve all objects in this region, with the additional constraint being that the parallax is between 0.69 and 4 mas. This latter constraint corresponds roughly to selecting stars at distances between 250 and 1450 pc from the Sun, that is, allowing any tails to be 200 pc closer to us than the cluster and up to 1 kpc farther. This arbitrary and very conservative query, which is validated a posteriori, resulted in 4 million objects. 

As we do not do any filtering on the data that we retrieve, we need to be careful when using the parallaxes to compute distances. Indeed, for stars fainter than $G$=18, the error bar on the parallax becomes larger than 0.2, which, given the distance of NGC 752, implies an error larger than 10\%. This, in turn, implies that we can no more do a simple inversion to determine the distance of the stars. One can, instead, use the photogeometric distances determined by \citet{2021AJ....161..147B}, which uses the parallax, colour and apparent magnitude of a star with a direction-dependent prior on distance, while exploiting the fact that stars of a given colour have a restricted range of probable absolute magnitudes (plus extinction). This is what we are using in the remaining of this paper.

\textit{Gaia} only provides radial velocities for objects brighter than around magnitude 13. Thus, for most objects, solely five astrometric parameters are available: positions (right ascension and declination), parallax ($\varpi$) and proper motion ($\mu_{\alpha^*},\mu_{\delta}$). Clustering in such a 5-dimensional space is generally useful to find cluster members \citep[e.g., ][]{Beccari2018}. However, this wouldn't be good enough to find out the extended tidal tails of a cluster. Even using only the 3-dimensional space of proper motions and parallax as done by \citet{2021MNRAS.505.1607B}  wouldn't guarantee finding the full extent of the tails, as these can be very much elongated in the radial direction, i.e., be at very different distances, but also have very different proper motions, as shown by Fig.~\ref{fig:petar}. It is thus necessary to rely on  a different methodology  \citep{Roeser+19,Oh2020,2021A&A...647A.137J}.

One way to deal with this is to use the convergent point method \citep[][]{Smart1939,Jos1999,vanLeeuwen09} that corrects for the fact that stars with different positions on the sky will have different proper motion values. 
The physical proper motions (in km/s, thus taking individual distances into account) of stars forming a spatially co-moving group point to a so-called convergent point (CP) on the sky. It is possible to compute the predicted parallel and perpendicular velocity components that point to the CP, $v_{||\mathrm{pred}}$ and $v_{\bot\mathrm{pred}}$, for a star with a given position in space or on the sky, by defining the CP such that $v_{\bot\mathrm{pred}}=0$. These values are then compared with the projected values of measured proper motions, $v_{||\mathrm{obs}}$ and $v_{\bot\mathrm{obs}}$. For stars that are co-moving, the differences $\Delta v_{||} = v_{||\mathrm{pred}} -  v_{||\mathrm{obs}}$ and $v_{\bot} \equiv v_{\bot\mathrm{obs}}$ will be close to zero. Stars from a given structure should also be connected spatially, so we also look at their Cartesian coordinates values in the Galactic frame, $u, v,$ and $w$.
    
We therefore computed for all stars from our \textit{Gaia} query, their $(u,v,w)$ as well as their $\Delta v_{||}$ and $v_{\bot}$, using the following cluster parameters\footnote{These were obtained after a first iteration of the method, using the values of the unambiguously detected members of the cluster itself.}, where we used as initial parameters the values provided in Simbad: 
\begin{align*}
[\mu^{*}_{\alpha},\mu_{\delta}, V_R] = 
[9.757 \, \mathrm{mas/yr}, -11.836\,  \mathrm{mas/yr}, 7.9 \, \mathrm{km/s} ], 
\end{align*}
and
\begin{align*}
[\mathrm{R.A.}, \mathrm{Dec}, \varpi] = [29.2009 \, \mathrm{deg}, 37.8066 \, \mathrm{deg}, 2.298 \,  \mathrm{mas}].
\end{align*}

We then used \textsc{dbscan}, in its scikit-learn implementation \citep{scikit-learn}, to find the largest cluster of points associated to NGC 752 in this 5-dimensional space. This was achieved using two different sets of \textsc{dbscan} parameters, namely $\epsilon=0.14$ and $n_{\rm sample}=9$, resulting in a sample of 538 stars, and $\epsilon=0.11$ and $n_{\rm sample}=2$, providing 586 stars, 102 of which were not in the previous sample. We therefore combined both samples, and are left with a final set of 640 stars, centred on the cluster, but showing nice, extended tails, covering about 40 degrees on sky, or about 300 pc at the distance of the cluster. This set of 640 stars is more than the double of the 282 members identified by \citet{2021MNRAS.505.1607B}. Our results are shown in Fig.~\ref{fig:tail}. 

One of the tails is slightly more populated than the other, but as shown in  Fig.~\ref{fig:tail}, they follow the expectations from the \textsc{petar} model, whether in right ascension and declination, or in galactic coordinates. Note that the distortion in the $u-v$ plane is a well-known effect due to the uncertainties on the parallaxes. The difference between the tails could be indicative of some additional interactions of NGC 752 with structures in the Galaxy. Detailed simulations will be needed to study this, and this is deferred to future work.

The lower left panel of Fig.~\ref{fig:tail} shows the \textit{Gaia} colour-magnitude diagramme (CMD), using the $G-R_p$ colour and the absolute $G$ magnitude. The position of each star in the CMD is corrected for its visual extinction, estimated using the 3D dust map  \textsc{Bayestar19}\footnote{\url{dustmaps.readthedocs.io}} 
\citep{2018JOSS....3..695M} -- giving a mean value of $A_v=0.2 \pm 0.1$. For comparison, a PARSEC isochrone\footnote{\url{http://stev.oapd.inaf.it/cgi-bin/cmd}} \citep{2012MNRAS.427..127B} for an age of 1.75 Gyr and a metallicity [Fe/H]=$-0.13$ is also shown. Despite their extremely wide extension on the sky, the  CMD obtained clearly resembles that of a simple stellar population down to an absolute magnitude $M_G=12$, with a well-defined turn-off and a red giant clump, as well as some white dwarfs and potential WD-main sequence binaries, all typical for an open cluster of the age of NGC 752. Only a handful of stars do not seem to follow the expectations, but we prefer not to remove them in an arbitrary way. Additional plots to help understand our sample are shown in Figs.~\ref{fig:A1}--\ref{fig:A6}.

We have made a selection of stars that belong to the central cluster, shown in orange in Fig.~\ref{fig:tail}. This corresponds to stars within 1.21 degrees -- the tidal radius estimated by  \citet{2021MNRAS.505.1607B}, corresponding to 9.4 pc at the distance of NGC 752.
This selection contains 320 stars and allows us to compute the mean parameters of the cluster:

\begin{align*}
\mathrm{R.A.} &=& 29.163375\pm0.467577\, \mathrm{deg},
\\
\mathrm{Dec} &=& 37.795161\pm0.357357 \, \mathrm{deg},
\\
\varpi &=& 2.281\pm0.166 \,  \mathrm{mas},
\\
\mu^{*}_{\alpha} &=& 9.77\pm0.51 \, \mathrm{mas/yr},  
\\
\mu_{\delta} &=& -11.78\pm0.54\,  \mathrm{mas/yr},
\\ 
V_r &=& 8.2\pm6.1 \, \mathrm{km/s} , 
\end{align*}
providing credence to the values we used above.
This translates in Galactocentric coordinates to\\
$[X, Y, Z]$ = $-8.294$, 0.275, $-0.158$ kpc, \\
$[V_x, V_y, V_z]$ = $-7.15$, 213.4, $-12.50$ km/s.

\begin{figure*}
	\includegraphics[width=18cm]{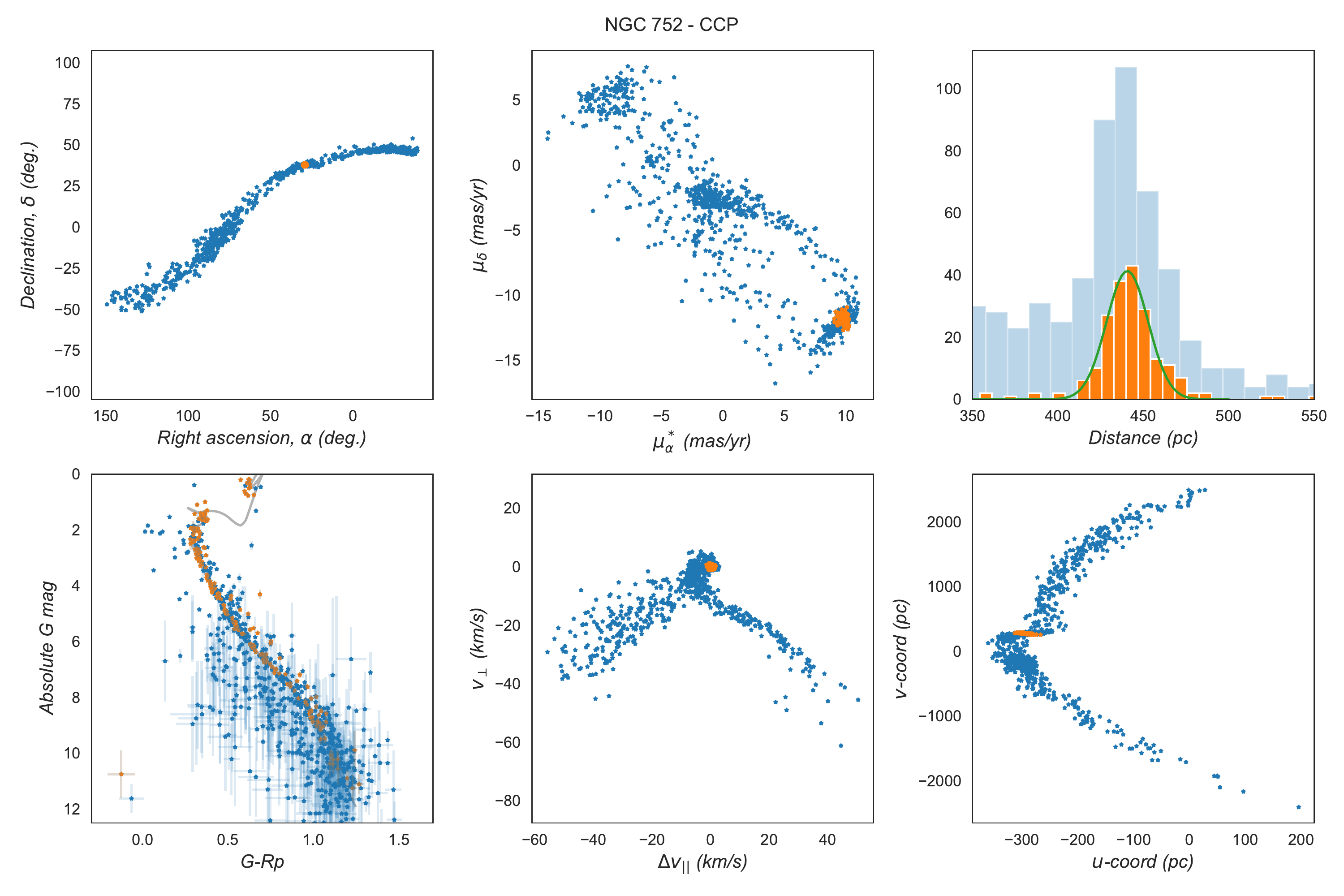}
    \caption{Same as Fig.~\ref{fig:tail} when applying the alternative CCP method as explained in Sec.~\ref{Sec:CCP}. Note the difference in scale. By construction, the selected stars follow the \textsc{petar} model. The distance histograms show only the most representative parts, as some stars have distances up to 3 kpc.}
    \label{fig:taillong}
\end{figure*}

As a check, we have verified that when using the criteria on proper motions and parallaxes applied by \citet{2021MNRAS.505.1607B}, we recover their sample (Fig.~\ref{fig:A1}). Removing the parallax constraint (as we expect the tails to be at different distances) leads to a sample of 375 stars, but does not allow to recover the full extent of the tails. This is further proof that this can only be done by using the CP method, as also the proper motions will depend on the positions. This is also shown by the fact that the stars selected by \citet{2021MNRAS.505.1607B} fill in a circular region with radius 2.2 km/s in the ($\Delta v_{||}, v_{\bot}$) plane. Using this criteria as done above leads to 387 stars that now cover the full extent of the tails. This is thus already an improvement. The outcome of the simulation shows, however, that one should not make such a selection in this ($\Delta v_{||}, v_{\bot}$)  plane. We look therefore in the next section, how this could be potentially improved.

\subsection{An alternative method}\label{Sec:CCP}
Figure~\ref{fig:petar} clearly shows that using only a small region either in proper motion or in convergent velocity spaces may be misleading. Indeed, the tails extend much more. To retrieve the full extent of the tails of the Hyades, \citet{2021A&A...647A.137J} improved on the convergent  point method, making use of the relation between the compact velocity, $\sqrt{v_\perp^2 + \Delta v_{||}^2}$, and the distance from the cluster in galactic coordinates, as seen in the numerical models. Such a relation for NGC 752 is shown in Fig.~\ref{fig:petar} and reveals a more complicated behaviour than in the case of the Hyades. The compact convergent point (CCP) method  introduced by \citet{2021A&A...647A.137J} can therefore not be followed exactly. Still, the observed relation could be used in an alternative way. 

It is clear, however, that we need to analyse a much larger area on sky than done previously. We therefore queried the \textit{Gaia} archive for all stars within a radius of 100 degrees around NGC 752 and with a parallax between 0.4 and 3. This led to a sample of 208 million stars. As such a sample is too large to be analysed, we performed several further cuts, driven by the relations that exist in the \textsc{petar} models of NGC 752. Thus, we defined polygons in the planes ($l,\mu_{\alpha}^*$), ($l,\mu_\delta$), and ($\varpi, \delta$), and used \textsc{shapely}\footnote{\url{https://shapely.readthedocs.io}} to select only those stars in the regions in interest (see Appendix~\ref{Appendix:CCPcuts}). The initial file has been divided into 96 subsamples, to allow parallel computing. All the resulting files amounted to 8 million  stars. 

With these remaining stars, we then looked at all the stars that were close  to the simulated stars in our \textsc{petar} model. More precisely, we used the metric introduced by \citet{Roeser+19} and kept only these stars that had
$$(u_*-u_{\rm m})^2 + (v_*-v_{\rm m})^2 + (w_*-w_{\rm m})^2 \leq r_{\rm lim}^2,$$
$$\frac{(\Delta v_{||,*}-\Delta v_{||,\rm m})^2}{a^2}+\frac{(v_{\perp,*}-v_{\perp,\rm m})^2}{b^2} \leq 1,
$$
where the symbol * indicates the stars in our reduced \textit{Gaia} catalogue, while $m$ is for the objects in our numerical simulations, and $r_{\rm lim}$=15 pc, $a=1.2$, and $b=0.5$. 
Such a selection allows to derive candidate stars belonging to the tails of NGC 752, over the whole region of interest. We find a total of 1041 stars, spanning the whole extent of the tails, as shown in Fig.~\ref{fig:taillong}.

The figure shows that one can, apparently, retrieve stars over the full extent of the tails -- several kpc from tip to tip! -- that share the same kinematics as those predicted for NGC 752 (see Fig.~\ref{fig:petar}). Similarly to the model, the stars so selected have distances from us between 350 and 3\,000 pc. The most distant stars will therefore be characterised by rather poor precision on their parameters, either their colours, but mostly their parallax. Using the photogeometric distances of \citet{2021AJ....161..147B} does improve the situation, but as seen in Fig.~\ref{fig:taillong}, there are still many stars for which the error on the absolute magnitude is rather large, making it difficult to ascertain that the selected stars belong to a single population. 

The present alternative method is rather conservative, given the very stringent parameters used, following in this \citet{Roeser+19}. This can be demonstrated by applying the same criteria to our \textsc{dbscan} sample. Of the original 640 stars in this sample, applying the above method, we are left with 341 stars, as seen in the third panel of Fig.~\ref{fig:A1} and, in more detail, in Fig.~\ref{fig:DBCCP}. It is important to note, however, that these 341 stars still do cover the full extent of the tails, so our conclusion about the length of the tails is not changed. Applying this alternative CCP method allows us to have a very pure sample -- as shown for example by the very nice CMD or the narrow distribution in distances (Fig.~\ref{fig:DBCCP}) -- but certainly not complete: the range in CP velocities (or in proper motions) is very limited as well. This likely highlights the fact that we are still dominated by errors from the \textit{Gaia} catalogue for the most distant and faintest objects. We have seen that if we wanted to retrieve the full \textsc{dbscan} sample with the alternative CCP method, we would need to use either  $a=9 , b=9, {\rm r_{lim}}=85$ pc, or $a=6 , b=4, {\rm r_{lim}}=80$ pc, which are likely more in line with the accuracy  on the \textit{Gaia} data. At the distance of NGC 752, a 10\% precision on the parallax indeed corresponds to an error of about 45~pc. 

Another caveat of the CCP method or the alternative one used here is that by design it assumes that the observations will follow the model. This may not  always be the case, if as seems to be found here, the tails are slightly asymmetric. 

Another unknown is the contamination rate among the selected stars from stars that would by chance have similar parameters to the cluster's members, including the tails. To assess this requires making detailed studies of mock catalogues, which is out of the scope of the current work. In any case -- and this is also valid for our \textsc{dbscan} sample -- one can only talk about candidate members, as a definitive proof would require a chemical analysis to ascertain the common origin of the stars. This may have to wait until the next generation of large scale spectroscopic surveys. 

The analysis presented in this section suggests that the CCP method proposed by our group or its alternative, supported with accurate dynamical models, will allow us to efficiently exploit future \textit{Gaia} data releases and find the full extent of tidal tails of open clusters. 

\begin{figure}
	\includegraphics[width=9cm]{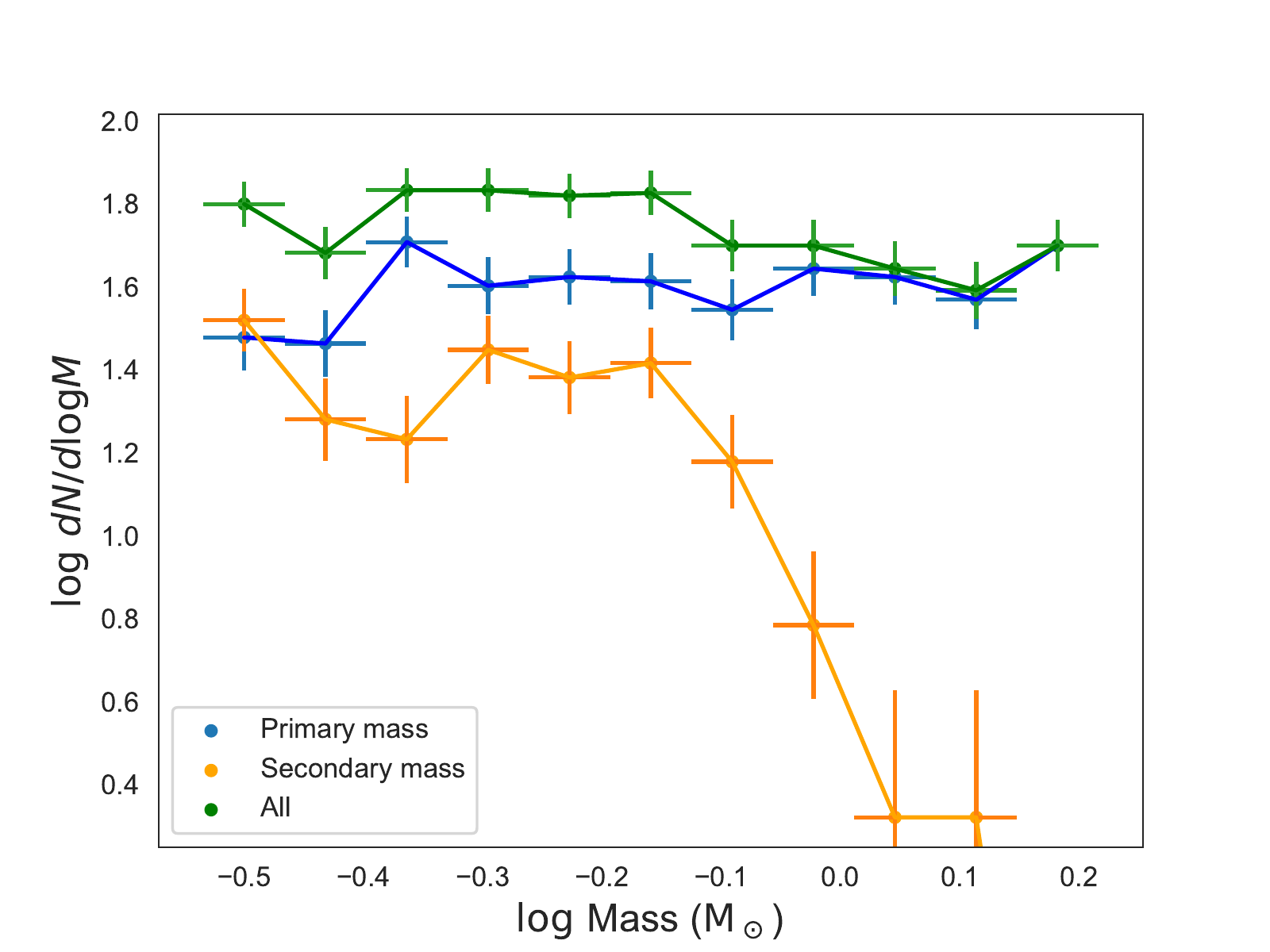}\\
	\includegraphics[width=9cm]{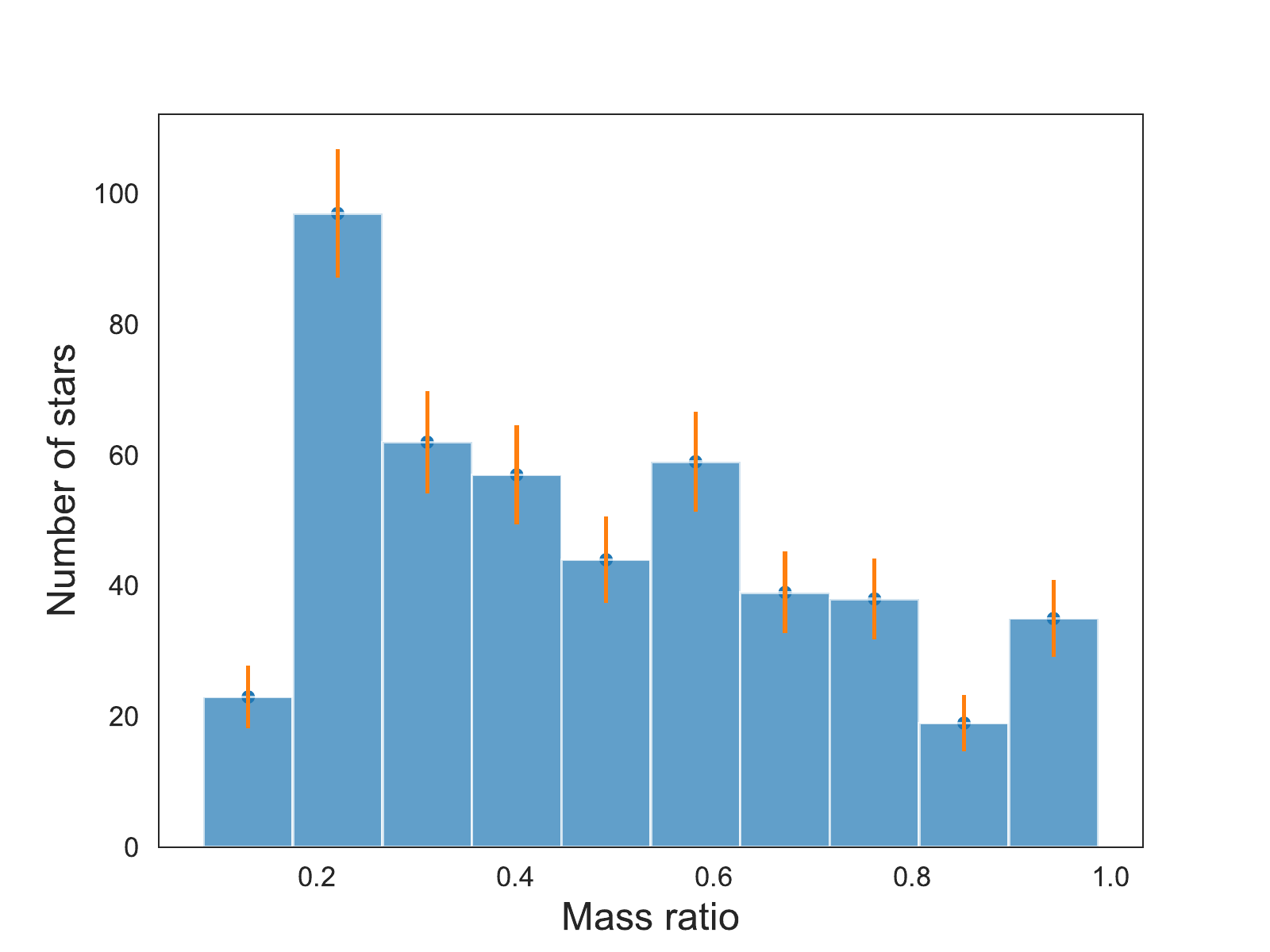}
    \caption{Top: The logarithm of the number of stars of a given interval in logarithm of the mass, for the \textsc{dbscan} sample. Indicated are the primary (or single) stars, the secondary stars, and the total mass of the systems. 
    Bottom: The mass-ratio distribution of our sample. As these mass ratios were determined photometrically, there is a bias against detecting binaries with a mass ratio below 0.2.}
    \label{fig:dbscan_mass}
\end{figure}

\begin{figure}
	\includegraphics[width=9cm]{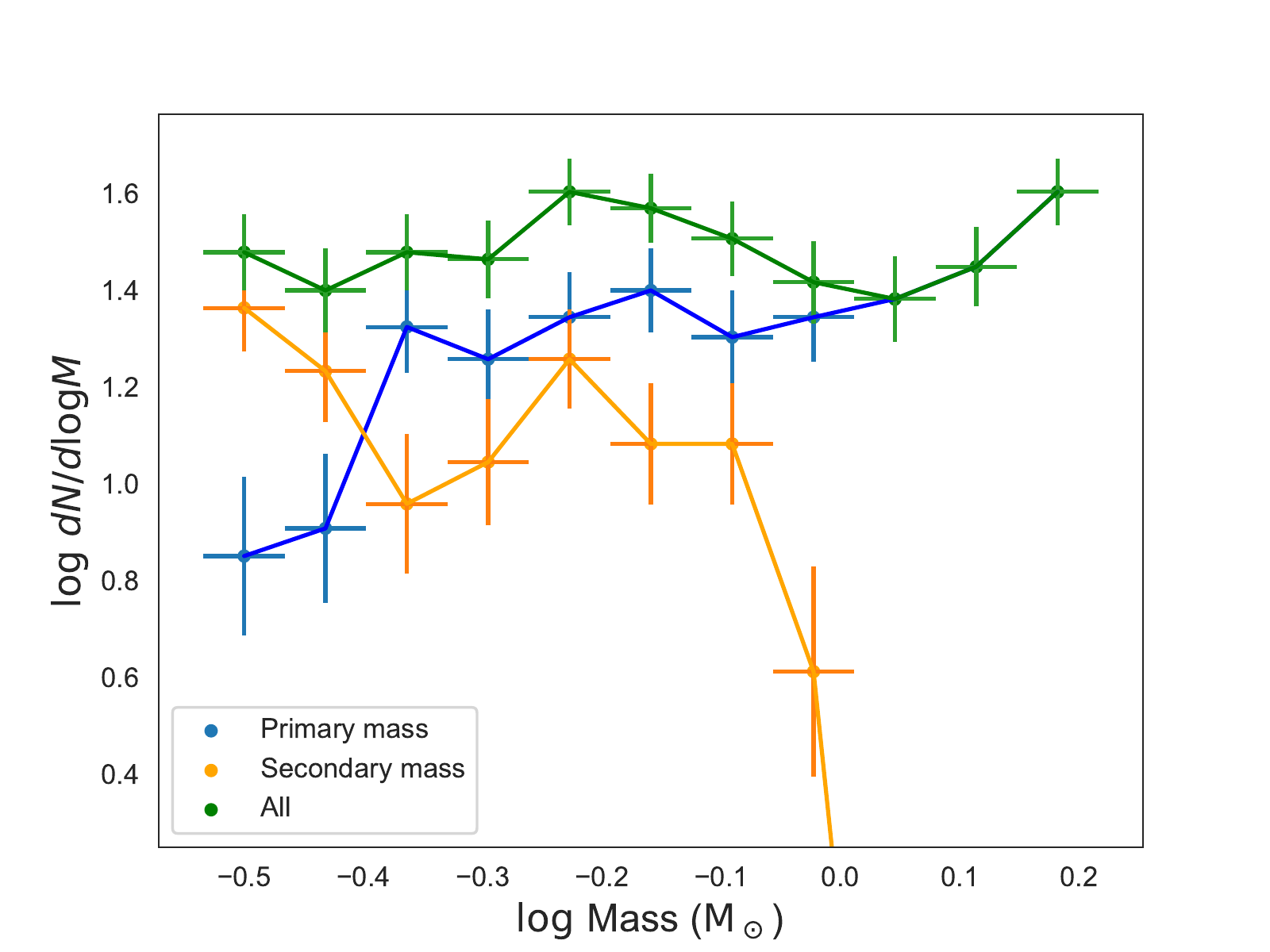}\\
	\includegraphics[width=9cm]{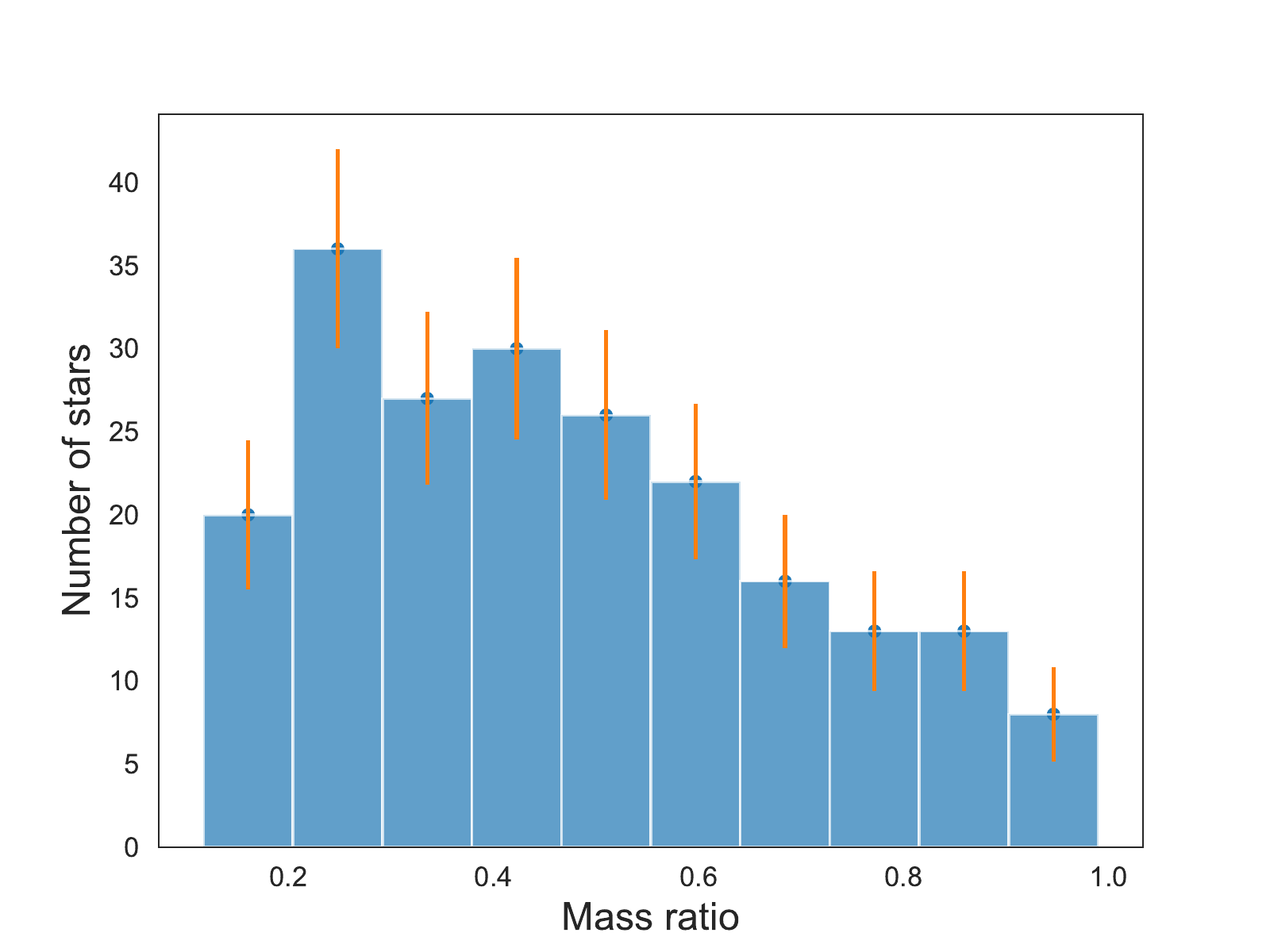}
    \caption{Same as Fig.~\ref{fig:dbscan_mass} for the stars belonging to the cluster only. Note the difference in scale compared to the previous figure.}
    \label{fig:clu_mass}
\end{figure}

\section{The stellar content}
\subsection{We shouldn't forget the binaries}
Using the CMD and comparing with  stellar isochrones, it is in principle possible to obtain an estimate of the masses of the stars, and thus of the various constituents of the whole structure. One should, however, beware that as the CMDs indicate in Figs.~\ref{fig:tail} and \ref{fig:taillong}, the selected samples are rich in binaries, and this needs to be considered, as it can potentially contribute to a non-negligible amount of mass, but mostly will lead to an increase in the number of low-mass stars, thereby modifying any slope of a stellar mass function  \citep{1991MNRAS.251..293K,Kroupa95c}. Thus, one can estimate the mass of the primary and the secondary using the mass-luminosity relations in the photometric bands of \textit{Gaia}. This is unfortunately degenerate, as it is not easy to know if a point in the CMD that is not on the main sequence corresponds to a brighter secondary (i.e., moving the system up on the CMD) or to a more massive primary and a redder companion (i.e., moving it to the right). We have taken a probabilistic approach to solve this issue: for each point, we consider all possible solutions in terms of binaries and then take the mean mass of the primary and secondary. 

Using photometry to derive binary masses has a severe caveat that only binaries with a mass ratio above 0.2 or so will be unambiguously detected, assuming that we can ignore any errors on the photometry or the parallax (which is not the case for the fainter stars). Moreover, NGC 752 contains also several red giants (18 in our \textsc{dbscan} sample), for which a companion can only be detected photometrically in the rare occurrence that the mass ratio is close to unity. \citet{1998A&A...339..423M} studied the binarity of 15 confirmed red giant members of NGC 752 and found four spectroscopic binaries, with orbital periods between 127 and 5276 days, with the companion being too faint to be detected directly. This is still a lower limit, as wider binaries could exist\footnote{The spatial resolution of \textit{Gaia} is about 0.2--0.4 arcsec, which at the distance of NGC 752 corresponds to about 85--170 au, much more than the maximum of about 10 au of the spectroscopic orbits.}. Thus, any total mass we derive should be considered a lower limit. Similarly, we did not consider in our computation the few white dwarfs, as well as the stars that lie below the main sequence -- possibly some binary white dwarfs, although we cannot exclude that these are main-sequence stars for which the extinction is larger than the value we used. As the total number of such stars is quite small in our \textsc{dbscan} sample, this will not have a big impact on the result.

The existence of the numerous red giants, together with the relatively large population of the upper main-sequence and the turn-off indicates that NGC 752 likely underwent mass segregation. However, determining the current mass function can only be done with a proper estimate of the secondary masses, as these will by definition always be lower than the primary ones. Thus, a  fraction of low-mass stars may be hidden in binary stars. 

\begin{figure*}
	\includegraphics[width=18cm]{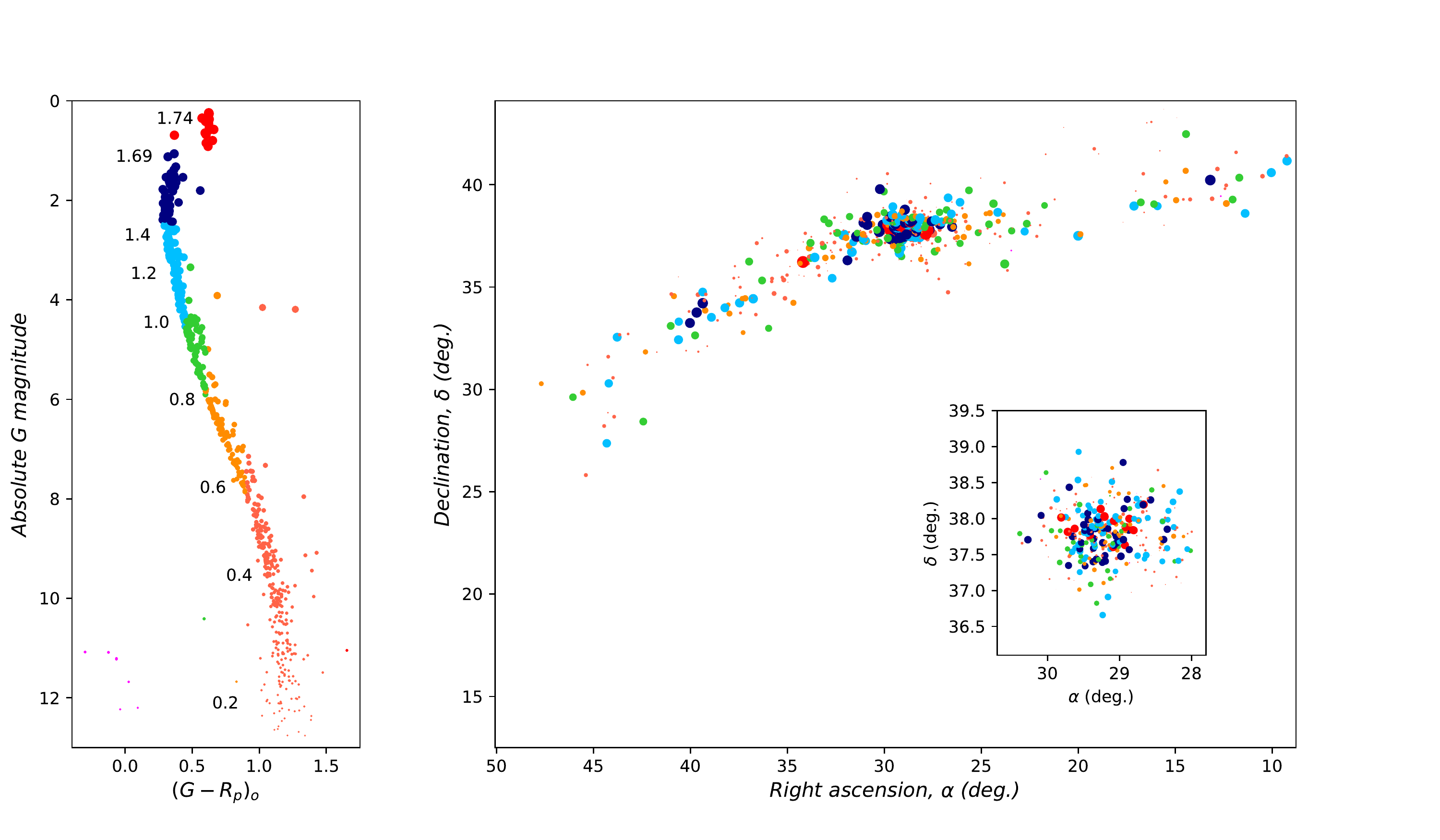}
	\caption{Another visualisation of the \textsc{dbscan} sample, to illustrate the mass distribution. The left panel shows the CMD, where points have colours according to their $(G-Rp)_o$ index or stellar evolution stage, and size according to their absolute magnitude in the $G$-band. The numbers correspond to the masses of the single stars in solar masses at the given location. The right panel shows the cluster and its tail as projected on the sky, with the same colour coding. The central part of the cluster is shown in the inset. A single colour version is offered in Fig.~\ref{fig:A8}.}
	\label{fig:tail2}
\end{figure*}

\subsection{The \textsc{dbscan} sample}

\subsubsection{Total mass}
To take into account the errors on the photometry and the parallaxes, we ran 10,000 Monte Carlo realisations, with values for the parameters extracted from a normal distribution around the mean and the standard error. This led to final values in our  sample of:
\begin{equation*}
  \begin{split}
  {\rm Binary~fraction:}&  ~0.69 \pm 0.01\\
 {\rm Total~cluster~mass:}& ~536 \pm 9~{\rm M}_\odot\\
 {\rm Total~mass~in~primaries~or~single~stars:}&  ~388  \pm 6~{\rm M}_\odot\\
 {\rm Total~mass~in~secondaries:}&  ~148 \pm 3~{\rm M}_\odot\\
\end{split}
 \end{equation*}

The binary fraction so determined may look slightly too high as it is expected that in field stars it will be only around 50\% for solar-type binaries, and even smaller for less massive stars. Although in clusters, this may be different -- that is, characterised by a larger fraction of binaries as we find here \citep[see also][]{Kroupa95c} -- if the fraction of binaries is actually smaller,  we are thus overestimating the fraction of binaries and thereby the amount of low-mass stars hidden as secondaries in our sample. This wouldn't affect much the total mass, but the mass function may not be entirely correct. 

Performing the same analysis on the central parts of the cluster, we find that it contains 65\% of binaries, with 237 M$_\odot$ in primaries, 81 M$_\odot$ in secondaries, for a total of 318 M$_\odot$. It is quite remarkable that these central parts are well fitted with a Plummer profile, with a total mass of 379 $\pm$ 21 M$_\odot$ and a Plummer scale of 0.53 $\pm$ 0.02 deg, which corresponds to 4.1 $\pm$ 0.2 pc. In this respect, we note that the Hyades \citep{Roeser+11}
and Praesepe \citep{Roeser+19} have also been shown to be well described by a Plummer model.

\subsubsection{Mass fraction}
We follow here \citet{2021MNRAS.505.1607B}: 
the present day mass function is calculated by binning the masses of the stars as obtained from the $M_G$ magnitudes into bins of width   $\delta \log M/M_\odot=0.068$. It is then possible to plot $\log dN/d\log M/M_\odot$.  
The results are shown in Fig.~\ref{fig:dbscan_mass}. The main difference to their work is that we do this for the primary stars (or mass of single stars), secondary stars, and for the sum of both primary and secondary masses. This shows that because of the numerous stars hidden in binaries, the number of low-mass stars is higher than what is found assuming the observed ``stars'' are all single. The lack of high-mass stars as secondaries is due to the fact that the primary of such stars will either be on the turn-off or are red giants, and such stars are very unlikely to be detected as binaries photometrically. The total mass distribution shows an increase of systems with lower masses, although the distribution is rather flat. We further note that as also highlighted by \citet{2021MNRAS.505.1607B}, because of the incompleteness of \textit{Gaia} data below $G \approx 20$, we also lack many primary and single stars with lower masses. 
Thus, one needs to be very careful when trying to derive the current mass function, using \textit{Gaia} data alone. We therefore refrain to make any quantitative analysis, such as trying to derive an exponent to the mass function. We stress, however, that our analysis points to an inverse slope to the one found by \citet{2021MNRAS.505.1607B}, and thereby highlights the need to be cautious. 

The bottom panel of Fig.~\ref{fig:dbscan_mass} shows the mass ratio distribution of the binaries in the \textsc{dbscan} sample. This seems to indicate a preference for systems with lower mass ratios, although a more careful study of the biases needs to be done before drawing any firm conclusions. 

The equivalent of Fig.~\ref{fig:dbscan_mass} for the inner parts of the cluster only, is shown in Fig.~\ref{fig:clu_mass}. This reveals perhaps an even flatter distribution of the primary mass, indicative of mass segregation. There is also an apparent excess of the most massive primaries. Given all the caveats presented earlier, we caution, however, in over-interpreting the data. 

\subsubsection{Minimum spanning tree}
That the cluster underwent mass segregation and that the tails mostly contain low-mass stars is further indicated in Fig.~\ref{fig:tail2}, which shows the stars in the tails and cluster, colour-coded according to their position in the CMD, a proxy for the stellar mass. The central parts of the cluster are clearly the richest in turn-off stars and red giants, the most massive of all for a single population -- that is, excepted for the invisible compact remnants.

Mass segregation can be quantified using the minimum spanning tree method~\citep[MST;][]{allis09}. The MST is the unique set of straight lines (``edges'') connecting a given sample of points (``vertices''; in this case the coordinates of the cluster' stars) without closed loops, such that the sum of the edge lengths is the minimum possible. Hence, the length of the MST is a measure of the compactness of a given sample of vertices (i.e., of a given population of stars). In particular, it has been proven in the literature that the ratio between the MST of a population of massive stars in a cluster -- taken as reference (MST$_{\rm ref}$) -- with respect to the MST of a sub-population of low-mass stars in the same cluster (MST$_{\rm low}$) provides a measurable indication of the degree of mass segregation in the given cluster~\citep[][]{ca04.348..589C}. 

We first derived the MST$_{\rm ref}$ of the massive stars in the absolute magnitude range $1.2 < M_G < 2.5$ and located in the area shown in the inset of Fig.~\ref{fig:tail2}. We count a total of 40 stars and derive MST$_{\rm ref}$=6.2 degrees. We then calculated the MST$_{\rm low}$ by randomly extracting 40 stars belonging to the cluster's population in the same region of the sky, but in the magnitude range $4.5 < M_G < 7.5$. For each extraction, we calculate the MST of the population. As shown in~\citet[][]{ca04.348..589C},
the MST calculated for the low-mass population follows a Gaussian distribution. In our case, the Gaussian fit of the distribution of MSTs is peaked
at 6.8$\pm0.3$ degrees, which we take as the value of MST$_{\rm low}$ and its associated uncertainty.
The degree of mass segregation is then evaluated as $\lambda=$MST$_{\rm ref}/$MST$_{\rm low}$=0.83$\pm$0.04. A value of $\lambda$ smaller than one indicates that the reference population (i.e., the massive stars) have a more compact distribution in the sky with respect to the low mass stars, as expected from the mass segregation.

\subsection{The CCP sample}
As mentioned previously, the larger sample cannot be analysed easily, as we would need to wait for the final \textit{Gaia} data release, to have a much better precision on the parallaxes of the stars in the most distant parts of the tails and the faintest ones. 
In the meantime, applying the above probabilistic approach, which takes into account the error bars on both magnitudes and parallaxes, we obtain that 576 M$_\odot$ are in single or primary stars, 241 M$_\odot$ in secondaries, for a total mass of 817 M$_\odot$.   
The equivalent of Fig.~\ref{fig:dbscan_mass} for the CCP sample is shown in Fig.~\ref{fig:A7}.

We note that the total mass of the stars in the cluster we found are about a factor two smaller than what we have in our \textsc{petar} simulations, which would indicate that the initial mass of the cluster (after the initial expulsion of gas; see the discussion in the introduction) was more like 2\,500 M$_\odot$, or slightly more when assuming that our selection is not complete. 
We therefore also ran a simulation of a cluster with initial mass 2\,500 M$_\odot$. However, such a cluster did not survive for as long as the age of NGC 752. Thus, the true result must lie in between.

\section{Lessons learned and forward look}
In this work, we have revisited the tidal tails of the open cluster NGC 752, using data from \textit{Gaia} EDR3. Numerical simulations using \textsc{petar} showed that such tails can be very long and present a very complicated kinematic signature, requiring an accurate statistical analysis that goes beyond the study of the proper motion and parallax hyperplane. Here, we prove  that one needs at the very least to use the convergent point method and the associated velocities. Numerical models are also required to verify that any possible tail does follow the expected trajectory, with any deviation from it being a possible indicator of a turbulent past or of the Galactic potential. 

In a first approach, we showed that using the CP method with the photogeometric distances, coupled with \textsc{dbscan} (with the best choice of parameters), we identified 640 candidate members of the cluster (which is well represented by a Plummer model) and its tails, more than doubling the previous estimate from \citet{2021MNRAS.505.1607B}. Although they can only be called candidates at this stage, they do seem to form a single structure and a single population. They span 260 pc on sky (from tip to tip), although they are slightly asymmetric, with the trailing tail being slightly less populated than the leading one -- the difference being not significant, however. They are thus almost four times as large as what was found by the previous authors. This indicates that when looking at tidal tails, it is necessary to explore a large region around the center of the target cluster and prefer the CP method. When analysing the stellar content of a cluster and its tails, it is also crucial to consider the binary population, as the majority of low-mass stars may well hide inside these systems. This isn't trivial to do with photometry only, however, and we will need future \textit{Gaia} data releases, which will include knowledge about binaries, as well as large spectroscopic surveys. 

We have also shown that by using numerical simulations and \textit{Gaia} data, it is in principle possible to find full-length tidal tails, spanning thousands of parsecs, allowing to better understand the past history of the clusters and how the population of field stars forms. This, however, requires a good understanding of possible contaminants that will only be gained by analysing in detail mock catalogues. Here also, only large spectroscopic surveys will allow us to confirm, by chemical tagging, the former members of the tails. By obtaining a drastic reduction in the number of candidates -- from hundred millions to a thousand stars -- the analysis we present here proves invaluable.  

Depending on the approach we take -- the limited one with \textsc{dbscan} or the extended one based on the CCP method -- the mass of the cluster is between about 40\% and 60\% of the total mass of the stars combined in the cluster and the tails. As about a third of the stellar mass of the cluster has been lost by stellar evolution, the initial mass of the cluster (that is, after any gas expulsion, which is not accounted for here) must have been roughly between 2 and 4 times the current one. The numerical simulations gives a factor around 8. Thus the initial cluster mass could be about 1,000 to 3,000 M$_\odot$,  although the upper limit is preferred as the lower mass clusters don't survive for 1.75 Gyr. This is still less than what is claimed by \citet{2021MNRAS.505.1607B}, who estimated that NGC 752 is a descendent of a Young Massive Cluster with initial mass of the order of 6,000--20,000 M$_\odot$. While these latter estimates rely on an analytical formula with unknown parameters, ours is based on direct comparison with numerical simulations and \textit{Gaia} data. Our  study is, however, unavoidably affected by the level of completeness of our algorithm, which is very hard to quantify. Nonetheless it is worth to stress the fact that in this work we identify a much larger number of cluster's members with respect to any of the past works.  We also give a complete new perspective to the size and properties of tidal tails in general.

In future work, we plan to address the various items mentioned above. We also aim at including primordial binaries in our $N$-body models, as \textsc{petar} can handle a large number of binaries.
Finally, we will further apply our methodology to a large set of open clusters, covering a wide range of ages, masses and locations in the Milky Way, in order to better constrain the formation and evolution of open clusters, and to understand the role they play in the evolution of the Galaxy.

\section*{Acknowledgements}

TJ acknowledges support through  the European Space Agency fellowship programme. LW is a JSPS International Research Fellow (Graduate School of Science, The University of Tokyo).
LW also thanks the support from the one-hundred-talent project of Sun Yat-sen University
and the National Natural Science Foundation of China through grant 12073090.
This research has made use of the
SIMBAD database and of the VizieR catalogue access tool, operated at CDS, Strasbourg, France. This work has made use of data from the European Space
Agency (ESA) mission \textit{Gaia} (\url{https://www.cosmos.esa.int/gaia}), processed by the \textit{Gaia} Data Processing and Analysis Consortium (DPAC, \url{https://www.cosmos.esa.int/web/gaia/dpac/consortium}). Funding for the
DPAC has been provided by national institutions, in particular the institutions
participating in the \textit{Gaia} Multilateral Agreement.

\section*{Data Availability}

The data used in this paper are available from the \textit{Gaia} EDR3 science archive, at \url{gea.esac.esa.int}.  
The $N$-body simulations were performed on the computing server of  L.W.
The data were generated with the code \textsc{petar}, which is available on \textsc{github}, at \url{https://github.com/lwang-astro/PeTar}.
The selected candidate members of the tails and the simulation data will be shared via  private communication with a reasonable request. 



\bibliographystyle{mnras}
\bibliography{NGC752} 



\appendix

\section{CCP pre-selection criteria}\label{Appendix:CCPcuts}

As mentioned in Sec.~\ref{Sec:CCP},  because the initial 100 degree-field around NGC 752 contained a sample of 208 million stars -- too large to be analysed as such -- we performed several further cuts, using the relations that existed in the \textsc{petar} models of NGC 752. Thus, we defined polygons in the planes ($l,\mu_{\alpha}^*$), ($l,\mu_\delta$), and ($\varpi, \delta$), and used \textsc{Shapely} to select only those stars in the regions in interest.
To allow readers to reproduce our results, we provide here the polygons used. For each couple of variables considered, we created two vectors, $x$ and $y$, that define a polygon and only keep all data points within these polygons.
The vectors are:
\begin{itemize}
    \item For $l$ (galactic longitude) -- $\mu_{\alpha}^*$: \\
x = [75.4,  121,  153,  283,   243, 132, 94.7]\\
y = [-2.9, 11.9, 11.9, -7.3, -16.9, 7.7, -4.9]
\item For $l$ -- $\mu_\delta$:\\
x = [85, 170, 223, 279, 266, 242, 171, 153, 99, 81]\\
y = [-4.3,- 18.1, -2.8, 3.5, 9.5, 9.2, -11.4, -11.9, 0.7, -2.8]
\item For $\varpi - \delta$: \\
x = [0.38, 2.69, 3.03, 2.9, 2.15, 0.27, 0.26, 2., 2.54, 2.44, 0.33]\\
y = [-57, -15, 0, 28, 46, 57, 40, 37, 19, -2, -45].
\end{itemize}

\section{Additional figures}

\begin{figure*}
	\includegraphics[width=8.5cm]{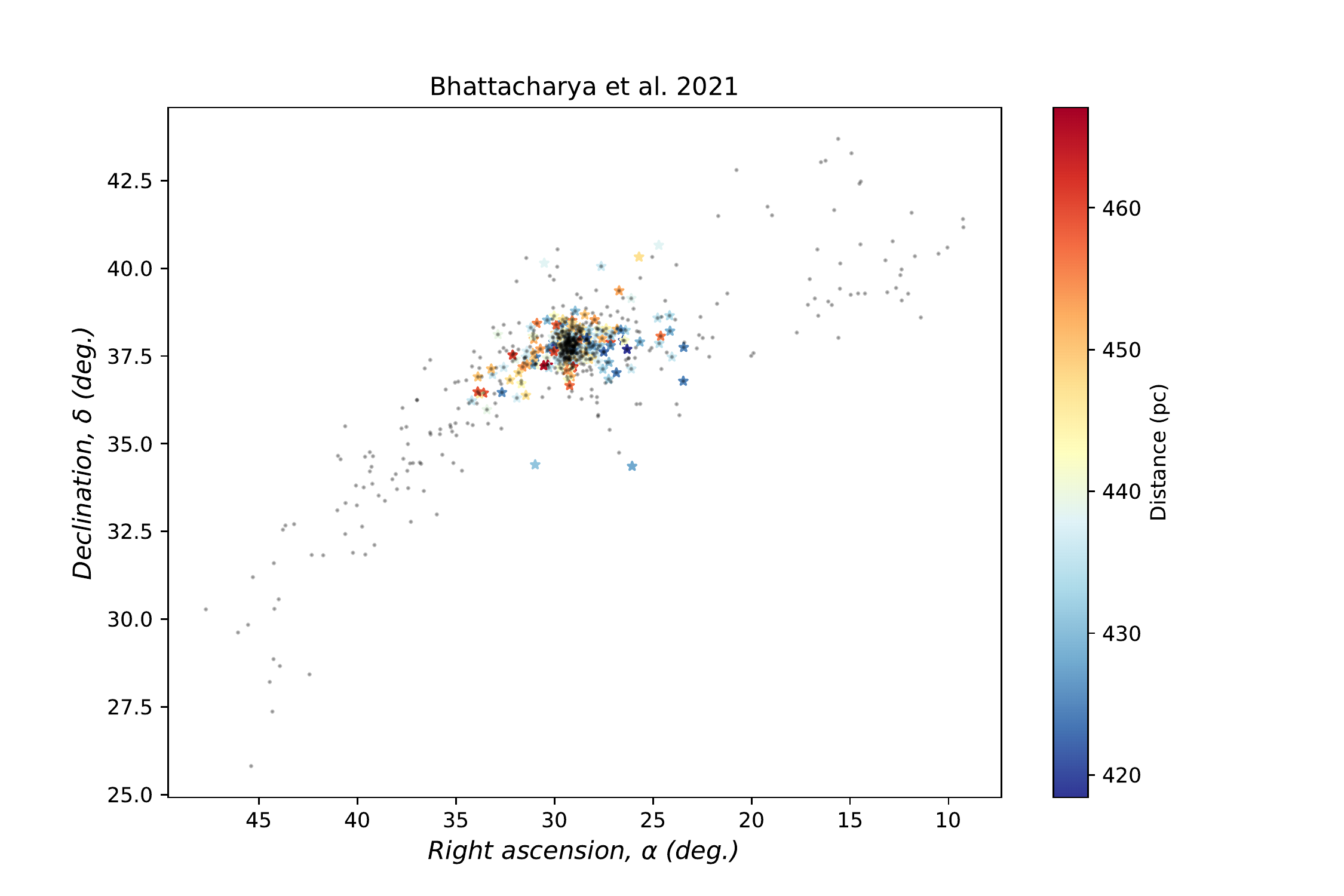}
	\includegraphics[width=8.5cm]{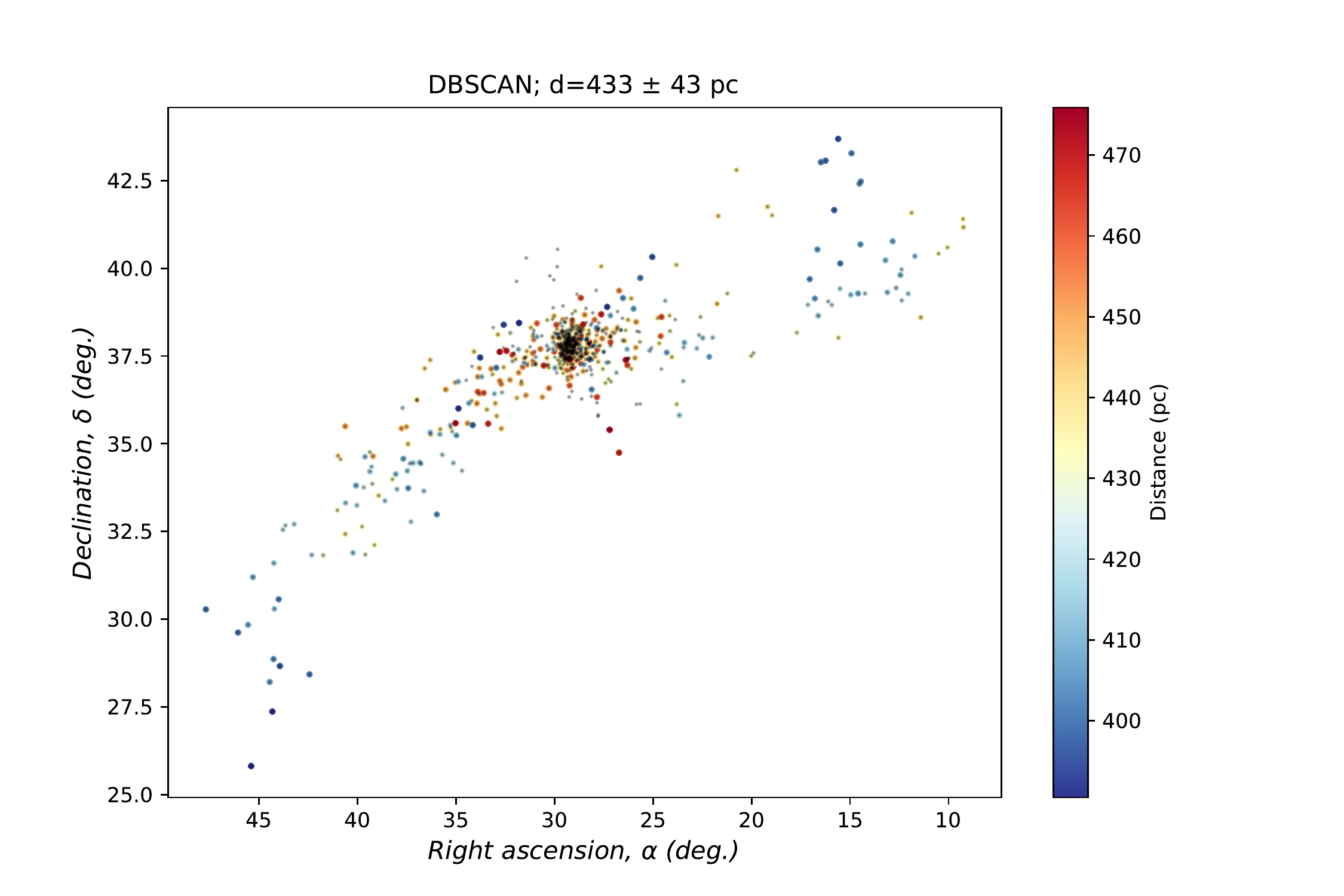}\\
	\includegraphics[width=8.5cm]{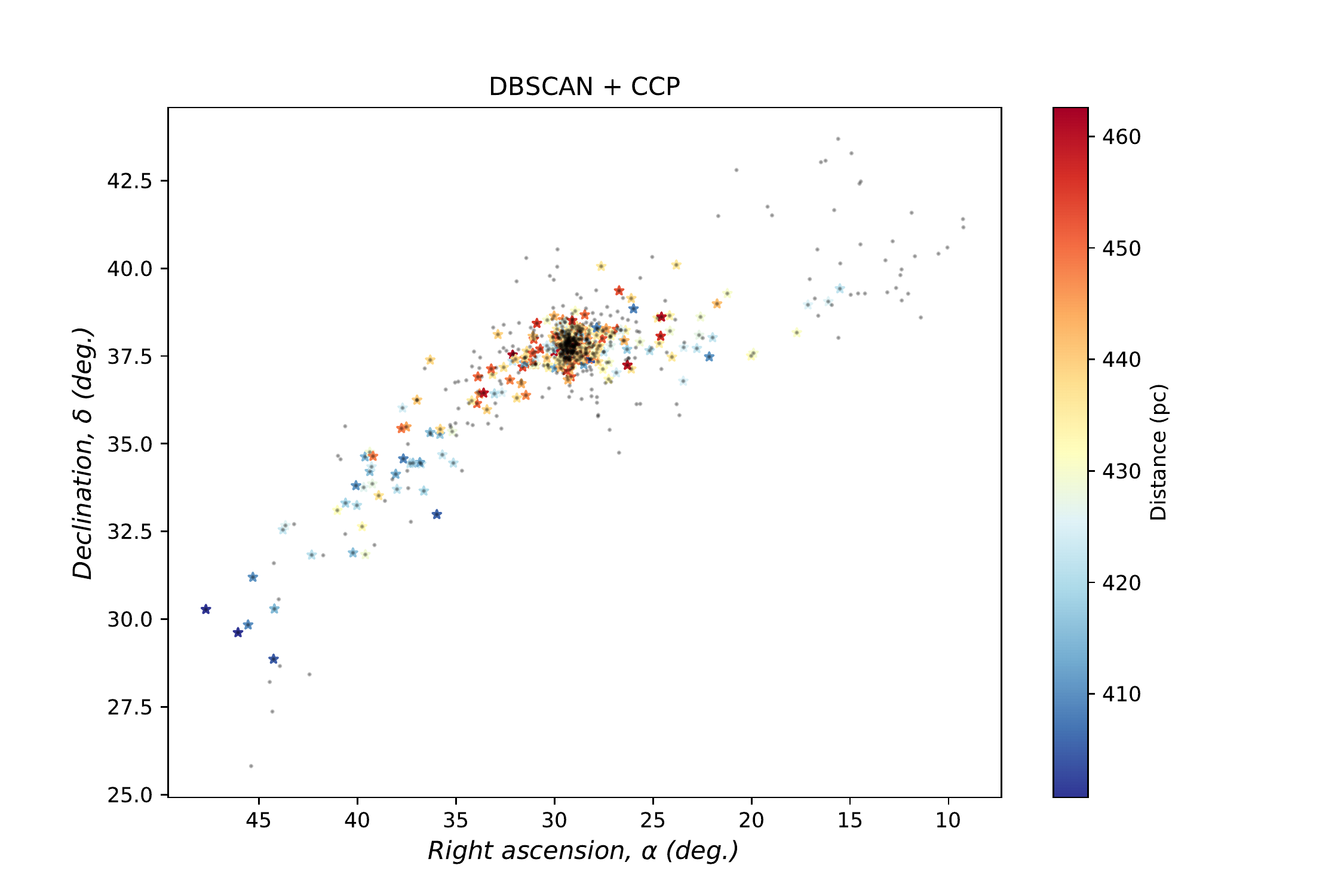}
	\includegraphics[width=8.5cm]{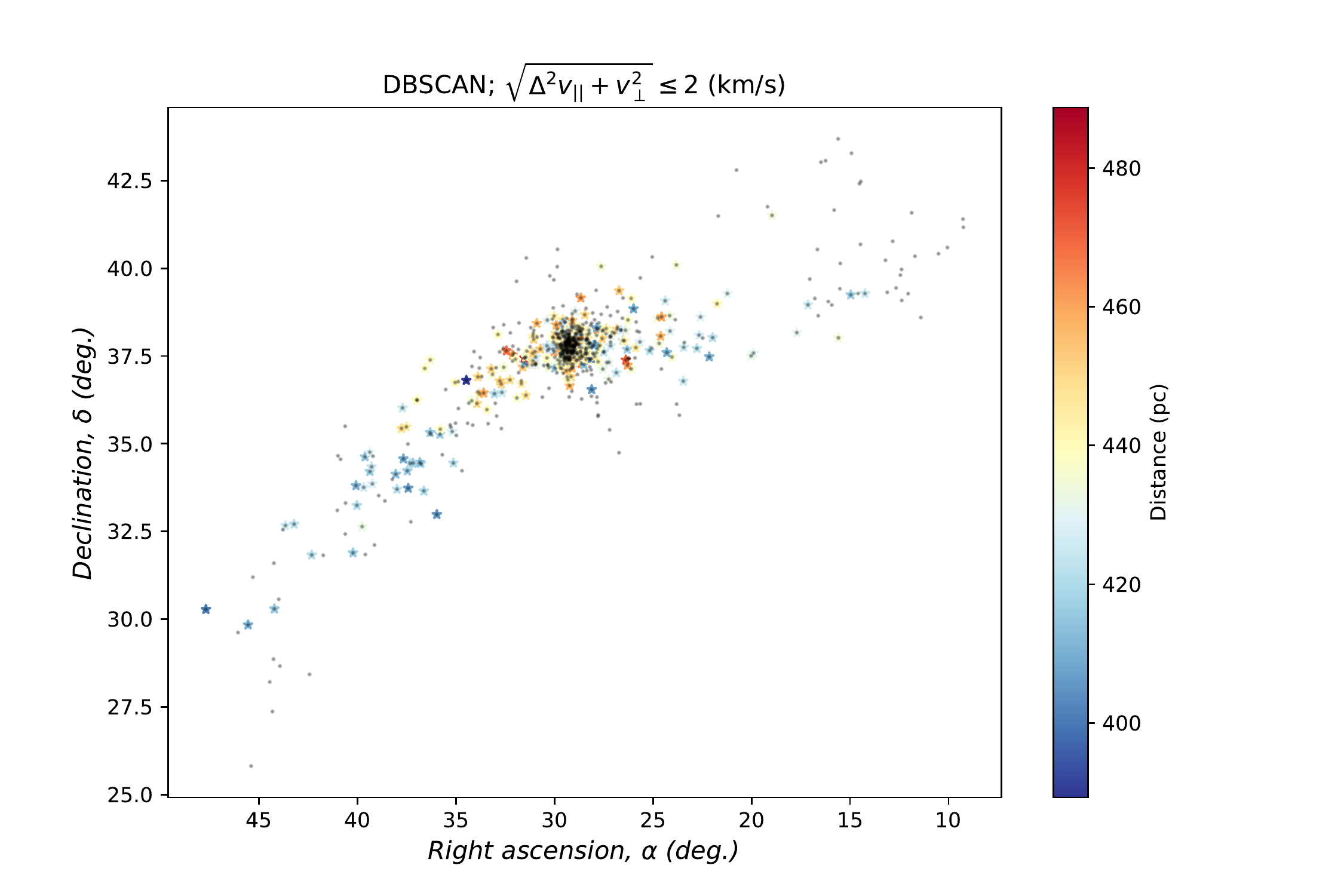}
	\caption{Stars selected when applying different cuts. From left to right and top to bottom: the selection of \citet{2021MNRAS.505.1607B}, indicating a much smaller extent of the detected tails; a selection of our \textsc{dbscan} sample in which only stars which are away from us in the range 433 $\pm$ 43 pc; applying the CCP method to our \textsc{dbscan} sample; and selecting only stars which have a CP velocity smaller or equal to 2 km/s. In all cases, the stars are coloured as a function of their distance to us, while the full \textsc{dbscan} sample we retrieve in this paper is shown as the background grey dots. It can be seen that in the last three cases, although there are less stars selected, the extent of the tails remains almost similar.}
	\label{fig:A1}
\end{figure*}

\begin{figure*}
	\includegraphics[width=8.5cm]{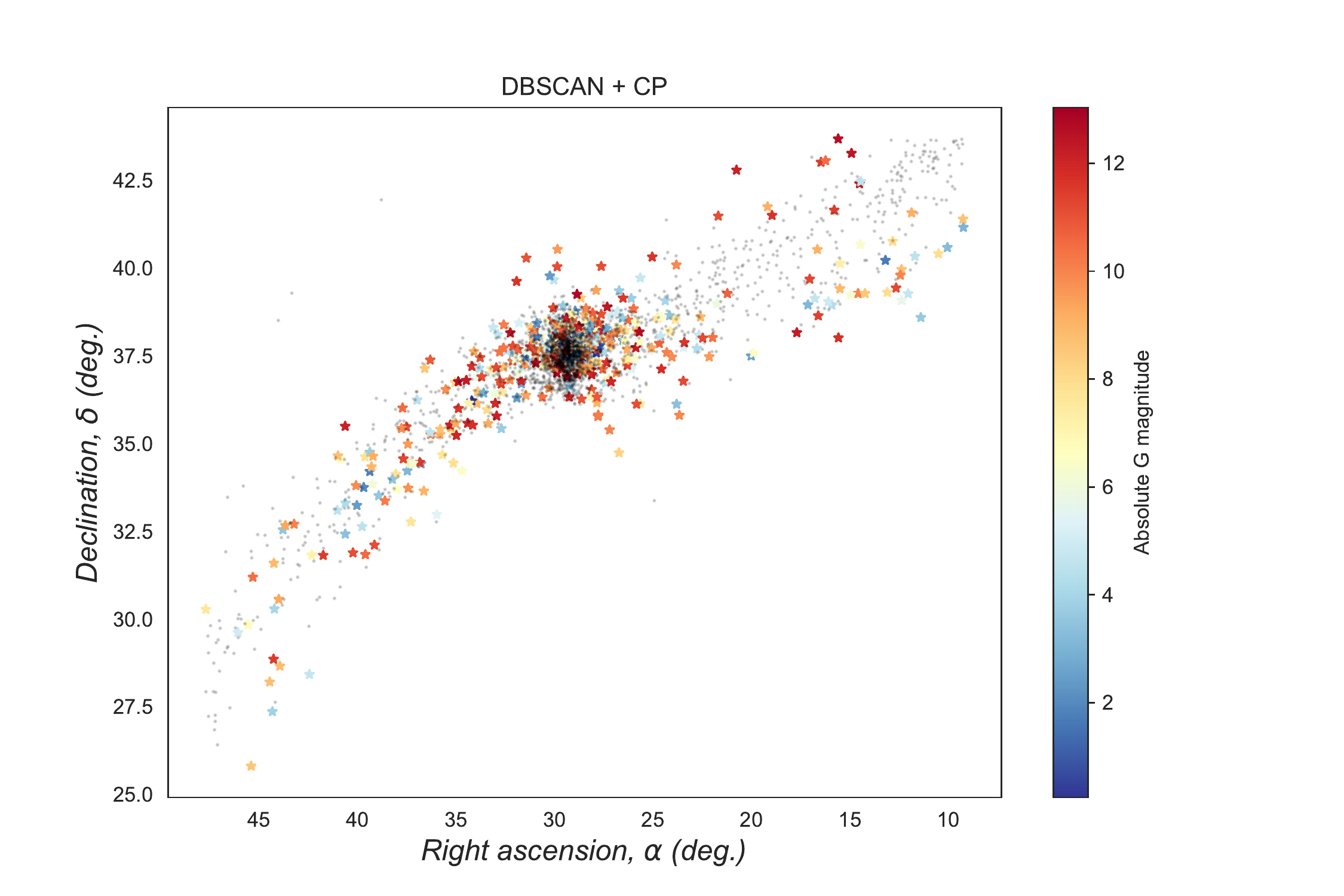}
	\includegraphics[width=8.5cm]{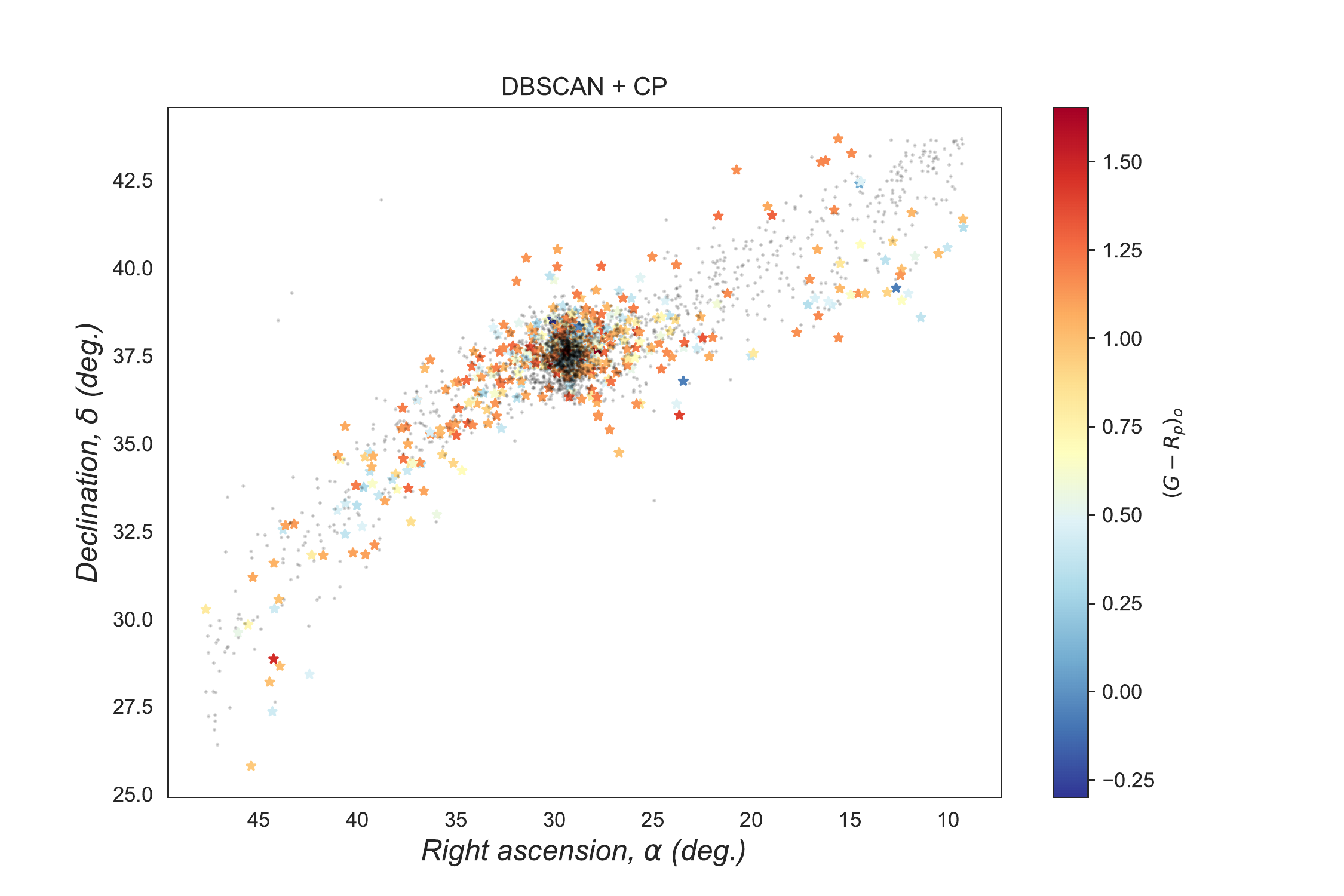}\\
	\includegraphics[width=8.5cm]{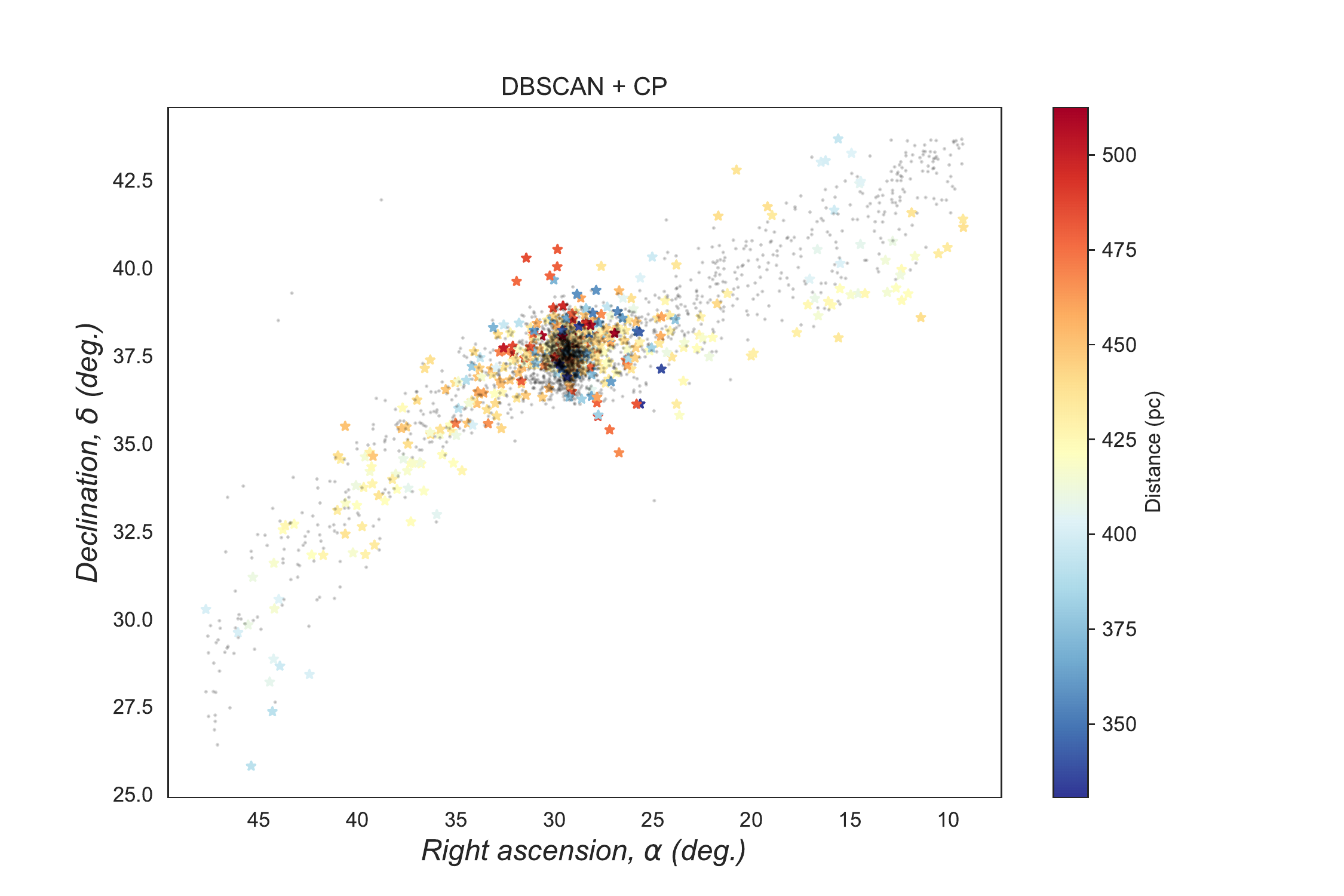}
	\includegraphics[width=8.5cm]{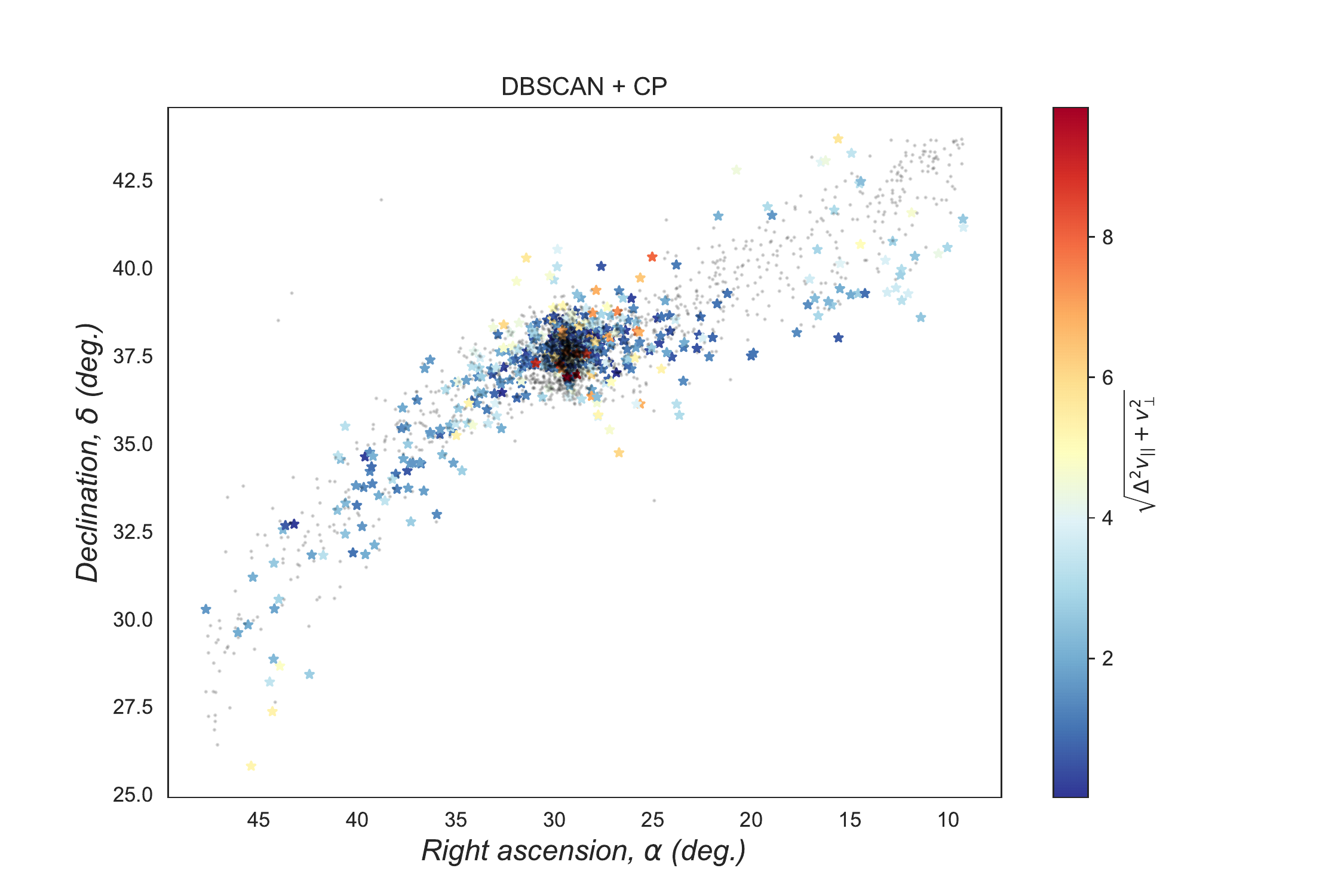}
	\caption{NGC 752 and its tidal tails as selected from \textit{Gaia} eDR3 data by \textsc{dbscan}. The plots show the same data but coloured according to a different variable. From left to right and top to bottom: the absolute, extinction corrected, \textit{Gaia} $G$-band magnitude; the \textit{Gaia} $G-R_p$ colour index; the distance; and the CP velocity.}
	\label{fig:A2}
\end{figure*}

\begin{figure*}
	\includegraphics[width=18cm]{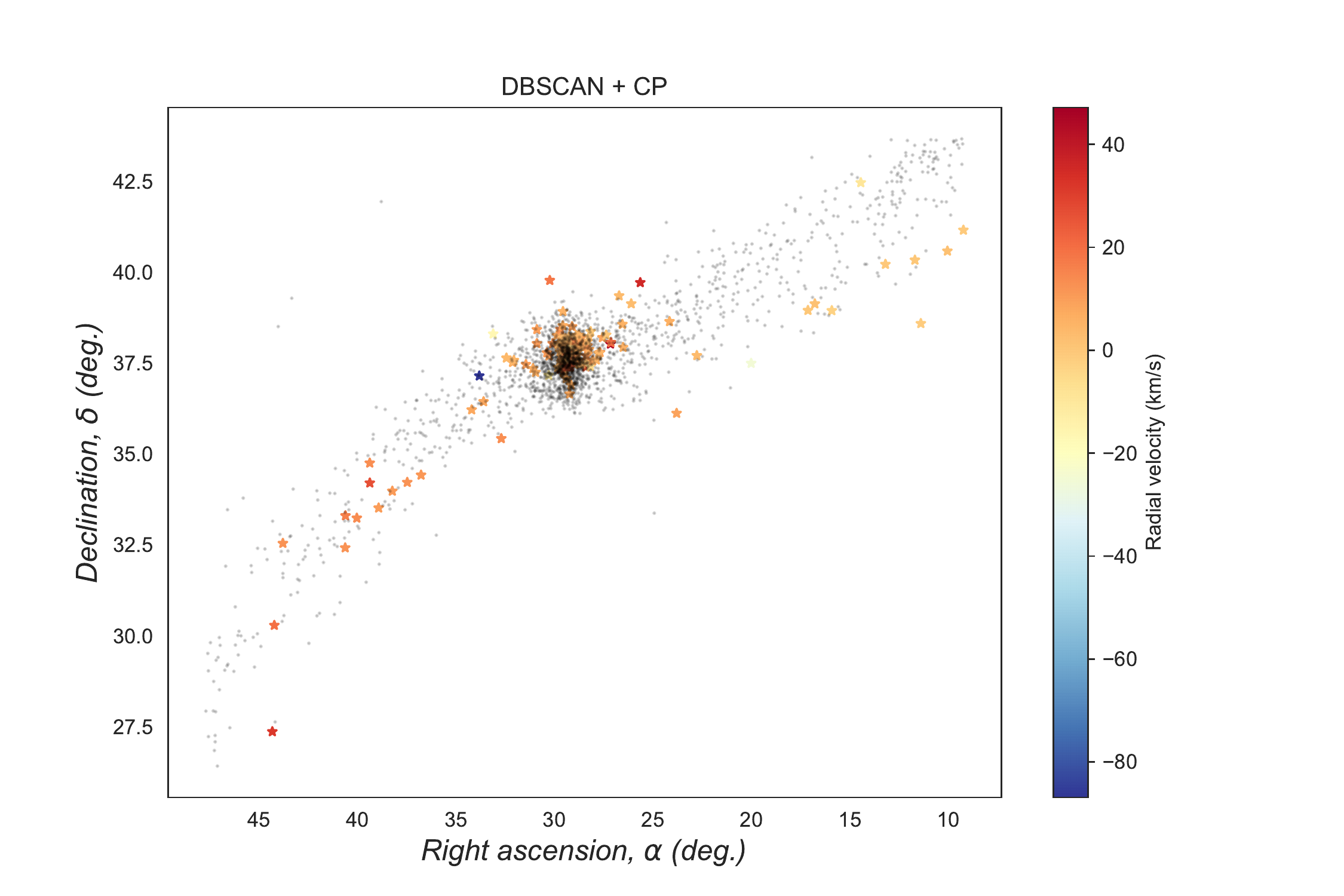}
	\caption{Same as Fig.~\ref{fig:A2} with the points coloured as a function of radial velocity. Only a subsample are shown -- those stars with radial velocities measured by \textit{Gaia}.}
	\label{fig:A6}
\end{figure*}

\begin{figure*}
	\includegraphics[width=18cm]{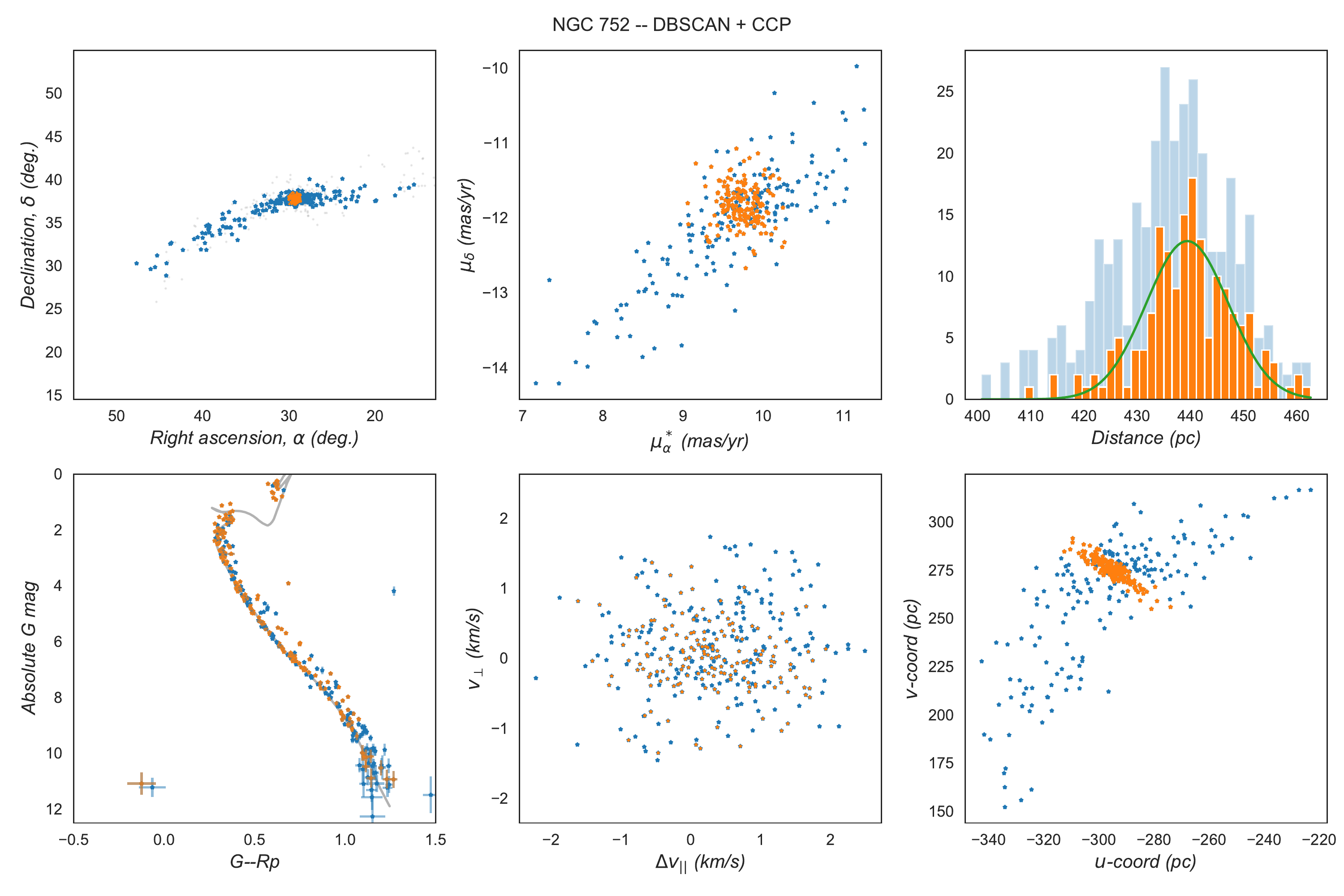}
	\caption{NGC 752 and its tidal tails as selected from \textit{Gaia} eDR3 data by \textsc{dbscan}, when applying the alternative CCP method -- this is thus a subset of our \textsc{dbscan} sample. The plots are the same as Fig.~\ref{fig:tail}, except that the background points are the full \textsc{dbscan} sample.} 
	\label{fig:DBCCP}
\end{figure*}

\begin{figure*}
	\includegraphics[width=9cm]{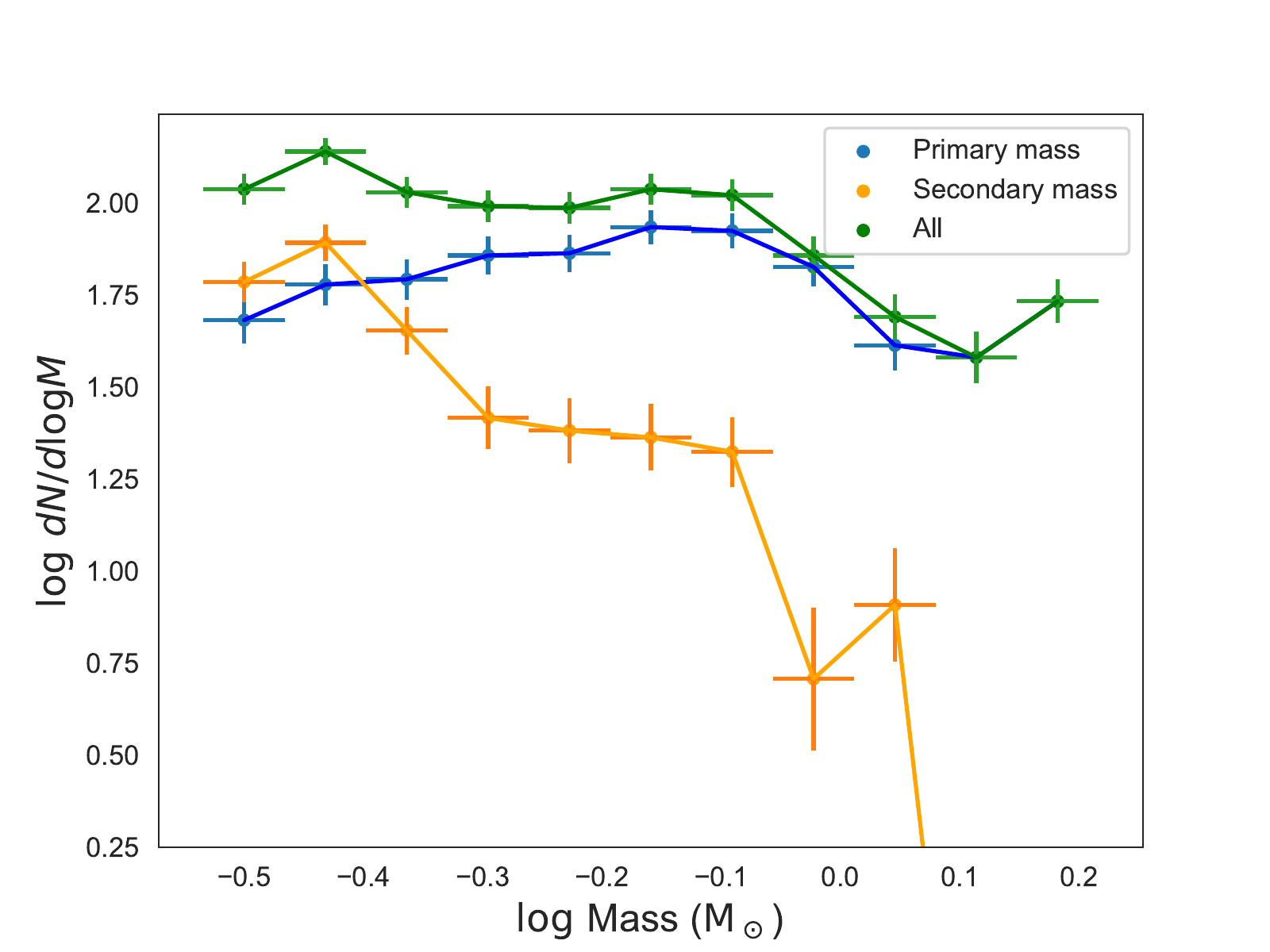}
	\includegraphics[width=9cm]{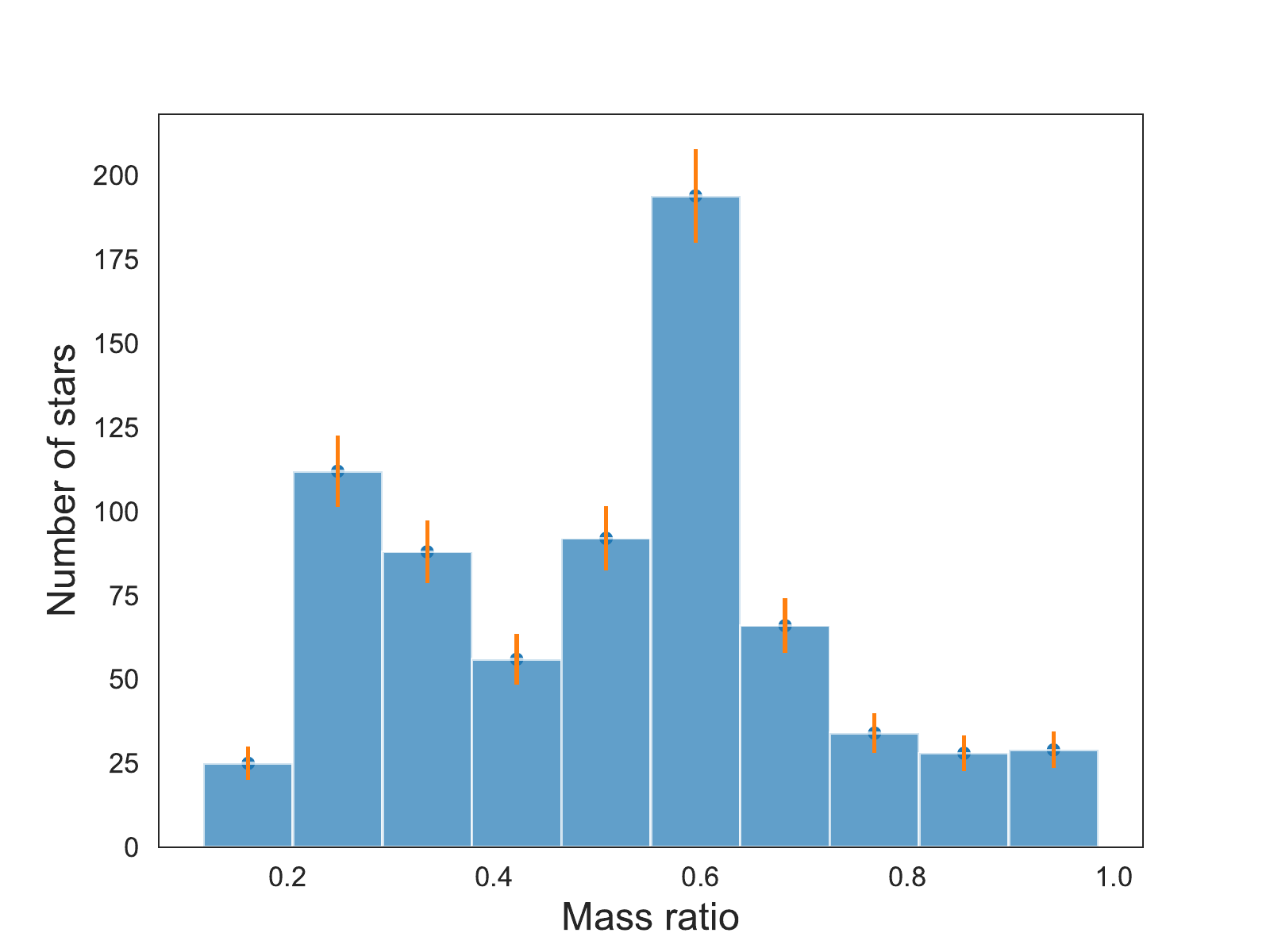}
    \caption{Same as Fig.~\ref{fig:dbscan_mass} for the stars selected by the alternative CCP method.}
    \label{fig:A7}
\end{figure*}

\begin{figure*}
	\includegraphics[width=18cm]{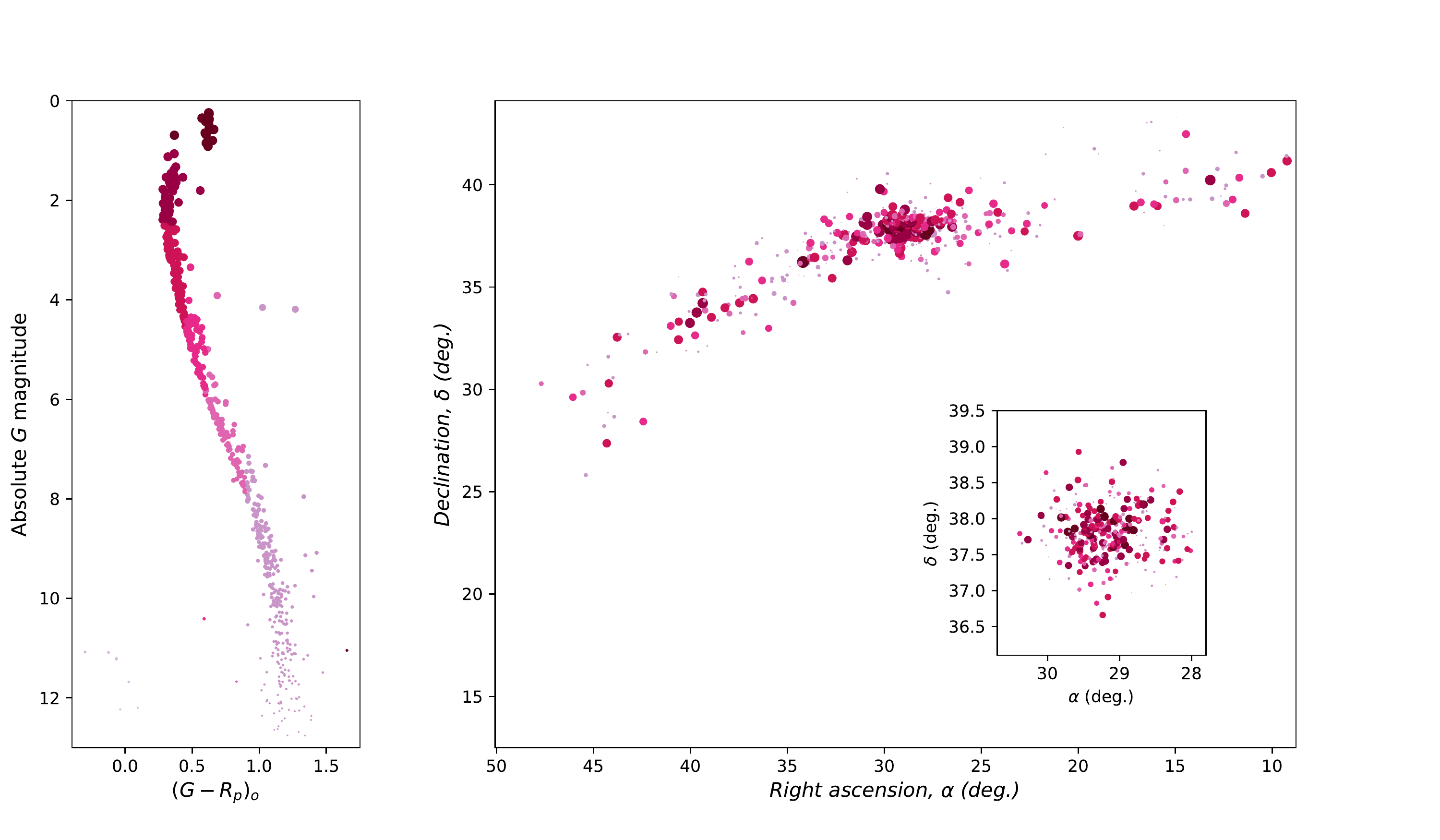}
    \caption{Same as Fig.~\ref{fig:tail2} where instead of colours, we use shades of the same colour to indicate the evolutionary state.}
    \label{fig:A8}
\end{figure*}

We provide here the additional Figs.~\ref{fig:A1}--\ref{fig:A8}, as referenced in the main text.


\bsp	
\label{lastpage}
\end{document}